\documentclass[12pt,english]{article}
\newcounter{withlatexpictures}
\usepackage{babel}
\usepackage{a4}
\usepackage{amsmath}
\usepackage{amssymb}
\usepackage{graphicx}
\usepackage[hang]{caption2}
\usepackage{subfigure}
\usepackage{fancyhdr}
\makeatletter
    \ifnum\@ptsize=2
    \fi
\makeatother
\ifnum\thewithlatexpictures=1
    \usepackage{feynmp}
\fi
\ifnum\thewithlatexpictures=1
    \newcommand{\feynmangraph}[4]{%
        \begin{fmffile}{#3}
            \begin{fmfgraph*}(#1,#2)
                #4
            \end{fmfgraph*}
        \end{fmffile}}
\else
    \newcommand{\feynmangraph}[4]{%
        \includegraphics[width=#1\unitlength]{#3.ps}}
\fi
\usepackage{hyperref}
\newcommand{\labeleq}[1]{\label{eq:#1}}
\newcommand{\labelfig}[1]{\label{fig:#1}}
\newcommand{\labelsec}[1]{\label{sec:#1}}
\newcommand{\refeq}[1]{(\ref{eq:#1})}
\newcommand{\reffig}[1]{\ref{fig:#1}}
\newcommand{\refsec}[1]{\ref{sec:#1}}
\newcommand{\massless}{``massless''}
\newcommand{\Bjorken}{Bjorken}
\newcommand{\MATHEMATICA}{\textsc{Mathematica}}
\newcommand{\TRACER}{\texttt{Tracer}}
\DeclareMathOperator{\tiD}{\mathnormal{D}}

\DeclareMathOperator{\Landau}{\mathcal{O}}
\DeclareMathOperator{\tir}{\mathnormal{r}}
\DeclareMathOperator{\trace}{Tr}
\DeclareMathOperator{\tibeta}{\beta}
\DeclareMathOperator{\kronecker}{\delta}
\DeclareMathOperator{\MOtheta}{\theta}

\DeclareMathOperator{\imagpart}{Im}

\newcommand{\DeclareMathFunction}[1]{\ensuremath{%
    \mathop{#1}}}
\newcommand{\diffn}[2][]{\DeclareMathFunction{%
  \mathrm{d}^{#1}{#2}}}
\newcommand{\comptont}[2][]{\DeclareMathFunction{%
    t_{#2}^{#1}}}
\newcommand{\comptonT}[2][]{\DeclareMathFunction{%
    T_{#2}^{#1}}}
\newcommand{\comptonttilde}[2][]{\DeclareMathFunction{%
    \tilde{t}_{#2}^{#1}}}
\newcommand{\comptonTtilde}[2][]{\DeclareMathFunction{%
    \widetilde{T}_{#2}^{#1}}}
\newcommand{\gpdf}[2][]{\DeclareMathFunction{%
    f_{#2}^{#1}}}
\newcommand{\gpdftilde}[2][]{\DeclareMathFunction{%
    \tilde{f}_{#2}^{#1}}}
\newcommand{\gpdH}[2][]{\DeclareMathFunction{%
    H_{#2}^{#1}}}
\newcommand{\gpdHtilde}[2][]{\DeclareMathFunction{%
    \widetilde{H}_{#2}^{#1}}}
\newcommand{\gpdE}[2][]{\DeclareMathFunction{%
    E_{#2}^{#1}}}
\newcommand{\gpdEtilde}[2][]{\DeclareMathFunction{%
    \widetilde{E}_{#2}^{#1}}}
\newcommand{\cfthm}[2][]{\DeclareMathFunction{%
    C_{#2}^{#1}}}
\newcommand{\ctildefthm}[2][]{\DeclareMathFunction{%
    \widetilde{C}_{#2}^{#1}}}
\newcommand{\negquad}{\ensuremath{\mspace{-18mu}}}
\newcommand{\detachedlhs}[1]{\ensuremath{\quad&\negquad{#1}\notag\\}}
\newcommand{\defeq}{\ensuremath{:=}}

\newcommand{\mcomma}{\ensuremath{\;,}}
\newcommand{\mlist}{\ensuremath{\mcomma\quad}}
\newcommand{\mperiod}{\ensuremath{\;.}}
\newcommand{\abs}[1]{\ensuremath{\lvert{#1}^{}\rvert}}
\newcommand{\lrabs}[1]{\ensuremath{\left\lvert{#1}\right\rvert}}
\newlength{\meanw}
\newlength{\meanwoffset}
\newcommand{\mean}[2][\meanwoffset]{\ensuremath{\text{%
    \settowidth{\meanw}{$\mathrm{#2}$}$#2$\hspace{-\meanw}%
    \makebox[\meanw]{\hspace{#1}$\overline{\phantom{\mathrm{#2}}}$}}}}
\newcommand{\cc}{\ensuremath{\ast}}

\newcommand{\bra}[1]{\ensuremath{\langle{#1}\lvert}}
\newcommand{\ket}[1]{\ensuremath{\rvert{#1}\rangle}}
\newcommand{\braket}[3]{\ensuremath{\bra{#1}{#2}\ket{#3}}}
\newcommand{\bordered}[2]{\ensuremath{\mathcal{#1}\left[{#2}\right]}}
\newcommand{\anti}[2][\meanwoffset]{\ensuremath{\mean[#1]{#2}}}
\newcommand{\cpveps}{\ensuremath{\varepsilon}}
\newcommand{\rpveps}{\ensuremath{\tilde{\varepsilon}}}
\newcommand{\inboson}[2][]{\ensuremath{%
    \cpveps_{#1}^{#2\vphantom{\mu}}}}
\newcommand{\inbosonp}[2][]{\ensuremath{%
    \cpveps'^{#2\vphantom{\mu}}_{#1}}}
\newcommand{\outboson}[2][]{\ensuremath{%
    \cpveps_{#1}^{#2\cc\vphantom{\mu}}}}
\newcommand{\outbosonp}[2][]{\ensuremath{%
    \cpveps'^{#2\cc\vphantom{\mu}}_{#1}}}

\newcommand{\inFermion}{\ensuremath{U}}

\newcommand{\outFermion}{\ensuremath{\anti{U}}}
\newlength{\diracw}
\newcommand{\dirac}[2][0pt]{\ensuremath{\text{%
    \settowidth{\diracw}{$#2$}$#2$\hspace{-\diracw}%
    \makebox[\diracw]{\hspace{#1}/}}}}
\newcommand{\diracp}[1]{\ensuremath{\dirac{#1}\vphantom{#1}'}}
\newcommand{\nullvec}{\ensuremath{\vec{\text{\textsl{0}}}}}

\newcommand{\dotprod}[2]{\ensuremath{{#1}{\cdot}{#2}}}
\newcommand{\pdotprod}[2]{\ensuremath{%
    (\dotprod{#1}{#2})}}
\newcommand{\intftn}[3][]{\ensuremath{%
    \int\frac{\diffn[#1]{#2}}{(2\pi)^{#1}}#3}}
\newcommand{\intfthm}[4]{\ensuremath{
    \int_{#2}^{#3}\frac{\diffn{#1}}{#1}#4}}
\newcommand{\intloop}[3][4]{\ensuremath{\intftn[#1]{#2}{#3}}}
\newcommand{\intn}[3][]{\ensuremath{\int{#3}\diffn[#1]{#2}}}

\newcommand{\alphaqcd}{\ensuremath{\alpha_{s}}}
\newcommand{\lambdaqcd}{\ensuremath{\varLambda_{\text{QCD}}}}
\newcommand{\MSbar}{\ensuremath{\overline{\text{MS}}}}
\newcommand{\CF}{\ensuremath{C_{F}}}
\newcommand{\TF}{\ensuremath{T_{F}}}
\newcommand{\ieps}{\ensuremath{\epsilon}}
\newcommand{\lceps}{\ensuremath{\varepsilon}}
\newcommand{\virtual}{\ensuremath{\ast}}
\newcommand{\elm}{\ensuremath{\text{em}}}
\newcommand{\xBjorken}{\ensuremath{x_{B}}}
\newcommand{\Mbar}{\ensuremath{\mean{M}\vphantom{M}}}
\newcommand{\pbar}{\ensuremath{\mean{p}\vphantom{p}}}
\newcommand{\Pbar}{\ensuremath{\mean{P}\vphantom{P}}}
\newcommand{\qbar}{\ensuremath{\mean[.2ex]{q}\vphantom{q}}}
\newcommand{\Qbar}{\ensuremath{\mean{Q}\vphantom{Q}}}
\newcommand{\shat}{\ensuremath{\hat{s}}}
\newcommand{\uhat}{\ensuremath{\hat{u}}}
\newcommand{\hq}{\ensuremath{h}}
\newcommand{\mhq}{\ensuremath{m_{\hq}}}
\newcommand{\etah}{\ensuremath{\eta}}
\newcommand{\etabarh}{\ensuremath{\mean{\eta}\vphantom{\eta}}}
\newcommand{\Ta}{\ensuremath{\text{A}}}
\newcommand{\Tl}{\ensuremath{\text{L}}}
\newcommand{\Tp}{\ensuremath{\text{P}}}
\newcommand{\Tt}{\ensuremath{\text{T}}}
\newcommand{\Ttflip}{\ensuremath{\overline{\text{T}}}}
\newcommand{\GeV}{\ensuremath{\text{GeV}}}
\begin{document}
%
%
\begin{titlepage}
  \title{
    \begin{flushright}\normalsize
      DO-TH 03/13\\
      December 2003
    \end{flushright}
    \begin{bfseries}
      Heavy quarks in deeply virtual Compton scattering
    \end{bfseries}
    \author{
      Jens~D.~Noritzsch%
      \thanks{Email address: \texttt{jens.noritzsch@udo.edu}}\\
      \begin{itshape}
        Institut f\"{u}r Physik, Universit\"{a}t Dortmund,
        44221 Dortmund, Germany
      \end{itshape}}}
    \date{}
\end{titlepage}
\maketitle
\thispagestyle{empty}
%
%
\begin{abstract}
  A detailed study of the heavy quark $h=c,b,\ldots$ contributions to
  deeply virtual Compton scattering is performed at both the amplitude
  and the cross section level, and their phenomenological relevance is
  discussed.  For this purpose I calculate the lowest order off-forward
  photon-gluon scattering amplitude with a massive quark loop and the
  corresponding hard scattering coefficients.  In a first numerical
  analysis these fixed order perturbation theory results are compared
  with the conventional intrinsic \massless\ parton approach considering
  generalized parton distributions for the heavy quarks.  The
  differences between these two prescriptions can be quite significant,
  especially at small skewedness where the massless approach largely
  overestimates the deeply virtual Compton scattering cross section.
\end{abstract}
%
%
\section{Introduction}
In deeply virtual Compton scattering
(DVCS)~\cite{Ji:1997ek,Radyushkin:1996nd,Ji:1997nm} a highly virtual
photon, generally radiated from a charged high-energy lepton, converts
to a real photon by scattering on a nucleon target that remains intact.
The phenomenology of DVCS looks very promising with first data in
different kinematic regions being available from several groups at fixed
target~\cite{Airapetian:2001yk,Stepanyan:2001sm} and collider
experiments~\cite{Adloff:2001cn,Chekanov:2003ya}.  On the theoretical
side the amount of uncertainties in the predictions for DVCS could be
reduced by incorporating next-to-leading order corrections in
perturbative quantum chromodynamics
(QCD)~\cite{Belitsky:2000gz,Freund:2001hm,Freund:2001hd} and power
corrections to the leading twist-two results starting at the twist-three
level~\cite{Kivel:2000fg,Belitsky:2001ns,Freund:2003qs}.

The main motivation for studying DVCS is the access to new
non-perturbative information on the structure of the nucleon carried by
generalized parton distributions
(GPDs)~\cite{Muller:1994fv,Ji:1997ek,Radyushkin:1996nd,Radyushkin:1996ru,Ji:1997nm,Collins:1997fb},
which will deliver insight into the spin structure of the nucleon, in
particular the quark and gluon spin and orbital angular momentum
contributions,~\cite{Ji:1997ek} and into the nucleon structure in three
dimensions~\cite{Burkardt:2000za}.  An excellent survey of the theory of
GPDs and their experimental accessibility can be found in a recent
comprehensive report~\cite{Diehl:2003ny}.

It is reasonable to inspect the inclusive proton structure function
$F_{2}$ in deep inelastic scattering (DIS) due to its close connection
to the general Compton amplitude via the optical theorem.  In the
kinematic region relevant for DVCS at the collider experiments the charm
contribution $F_{2}^{c}$ to the proton structure function has been found
to be up to $20\%$~\cite{Adloff:2001zj,Chekanov:2003rb}.  If one
presumes this same ratio for the corresponding form factor of the DVCS
amplitude this results in a charm contribution of about one third to the
DVCS cross section (amplitude squared).  Hence a proper treatment of
charm in DVCS can be quite important.

Generally, heavy quarks $h=c,b,t$ are characterized by
$\mhq^{2}\gg\lambdaqcd^{2}$ whereas light (\massless) quarks have
$m_{u,d,s}^{2}\ll\lambdaqcd^{2}$.  In fixed order (FO) perturbation
theory, heavy quarks contribute in lowest order to the photon-gluon
scattering (PGS) $\gamma^{\virtual}g\to\gamma{}g$ DVCS subprocess where
they appear as massive internal fermion lines.  As is well known, such a
fixed order calculation correctly describes the threshold region
$\shat\gtrsim4\mhq^{2}$ in
DIS~\cite{Adloff:2001zj,Chekanov:2003rb,Gluck:1998xa,Harris:1995tu},
and its perturbative stability has been demonstrated in
Ref.~\cite{Gluck:1994dp} even up to $\shat\sim10^{6}\,\GeV^{2}$.  In the
conventional parton model approach potentially large logarithms
$\log^{n}(\mhq^{2}/\shat)$ appearing in the fixed order PGS result are
resummed to all orders and absorbed into intrinsic \massless\ parton
(MP) distributions for the heavy flavors~\cite{Collins:1986mp}.

In the next section, after giving a rough overview of DVCS in the
leading twist approximation, the technicalities of incorporating heavy
quark contributions in the DVCS amplitude are described and the need for
massive hard scattering coefficient functions is explained.  In
Sec.~\refsec{pgs} a systematic way to calculate the lowest-order of PGS
with a massive quark loop is introduced whose relevant results are
presented in Sec.~\refsec{analytics}.  This completes the requisites for
a first numerical analysis of heavy quark effects in DVCS where the
numerical results at the amplitude as well as at the cross section level
are discussed in Sec.~\refsec{numerics}.  Finally, in
Sec.~\refsec{theend} the conclusions are drawn and an outlook is given.
%
%
\section{Leading twist DVCS amplitude}
First, some general properties of DVCS have to be reiterated where
mainly the notations of Refs.~\cite{Ji:1998xh, Belitsky:2000gz} are
used.  The general kinematics of DVCS on a proton are shown in
Fig.~\reffig{kinematics}
\begin{figure}
  \centering
  \feynmangraph{30}{24}{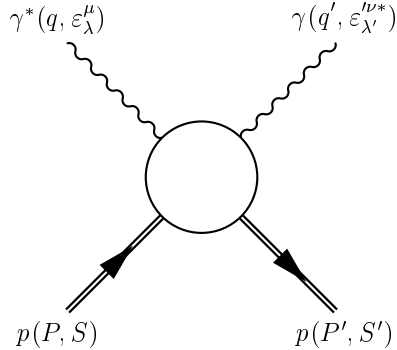}{
    \fmfleft{iN,ip}
      \fmflabel{\makebox[0pt]{%
        $\gamma^{\virtual}(q,\inboson[\lambda]{\mu})$}}{ip}
      \fmflabel{\makebox[0pt]{$p(P,S)$}}{iN}
    \fmfright{oN,op}
      \fmflabel{\makebox[0pt]{%
        $\gamma(q'^{\vphantom{\mu}}_{\vphantom{\lambda'}},%
        \outbosonp[\lambda']{\nu})$}}{op}
      \fmflabel{\makebox[0pt]{$p(P',S')$}}{oN}
    \fmfpen{thin}
      \fmf{double_arrow}{iN,v}
      \fmf{photon}{ip,v}
      \fmfv{decor.shape=circle,decor.filled=empty,decor.size=.333w}{v}
      \fmf{double_arrow}{v,oN}
      \fmf{photon}{v,op}
  }
  \caption{General kinematics of deeply virtual Compton scattering on a
    proton.}
  \labelfig{kinematics}
\end{figure}
with the masses defined to be $Q^{2}\defeq-q^{2}$ and
$M^{2}\defeq{}P^{2}=P'^{2}$.  An independent set of momenta is given by
\begin{equation}
  \Pbar\defeq
    (P+P')/2\mlist
  \qbar\defeq
    (q+q')/2\mlist
  \varDelta\defeq
    P'-P
\end{equation}
that leads to the following five Lorentz invariants
\begin{equation}
  \Qbar^{2}\defeq
    -\qbar^{2}\mlist
  x\defeq
    \frac{\Qbar^{2}}{2\dotprod{\Pbar}{\qbar}}\mlist
  \xi\defeq
    -\frac{\dotprod{\varDelta}{\qbar}}{2\dotprod{\Pbar}{\qbar}}\mlist
  t\defeq
    \varDelta^{2}\mlist
  \Mbar^{2}\defeq
    \Pbar^{2}=M^{2}-t/4
\end{equation}
where for the moment the outgoing photon is also considered virtual.
Only the first three invariants survive the generalized \Bjorken\ limit
$\Qbar^{2}\to\infty$ with $x,\xi$ fixed in which the first two are
straightforward generalizations of corresponding invariants in DIS.  The
additional scaling variable $\xi$, the so-called skewedness, is a
measure of the difference between the two photon virtualities.
Furthermore, the following invariants prove to be helpful for the
representation of the analytical results in Sec.~\refsec{analytics}
\begin{gather}
  \shat\defeq
    (\Pbar+\qbar)^{2}=\Qbar^{2}(-1+1/x)+\Mbar^{2}\mlist
  \uhat\defeq
    (\Pbar-\qbar)^{2}=\Qbar^{2}(-1-1/x)+\Mbar^{2}\mcomma\notag\\
  q^{2}=
    (\qbar+\varDelta/2)^{2}=\Qbar^{2}(-1-\xi/x)+t/4\mlist
  q'^{2}=
    (\qbar-\varDelta/2)^{2}=\Qbar^{2}(-1+\xi/x)+t/4
\end{gather}
where $\Mbar^{2}$ and $t/4$ are neglected in the generalized Bjorken
limit.  In DVCS with a real photon in the final state only one scaling
variable remains
\begin{equation}
  x\approx\xi=x(1-t/4\Qbar^{2})\mlist
  \Qbar^{2}=Q^{2}/2+t/4\approx{}Q^{2}/2\mcomma
\end{equation}
which is chosen to be $\xi$ to avoid confusion with the usual \Bjorken\ 
variable $\xBjorken\approx2\xi/(1+\xi)$ in DIS, and it is more
appropriate to use $Q^{2}$.

The dynamics is contained in the DVCS amplitude
\begin{equation}
  -i\comptonT[\nu\mu]{}
  =\intn[4]{z}{e^{i\dotprod{\qbar}{z}}
    \braket{P'S'}{\bordered{T}{J^{\nu}(z/2)J^{\mu}(-z/2)}}{PS}}
\end{equation}
that is given by the time-ordered product of two currents which are
related to the electromagnetic current by
$J_{\elm}^{\mu}(z)=eJ^{\mu}(z)$.  The leading twist contributions, apart
from those that arise due to photon helicity flip which are beyond the
scope of this work, are contained in the following two, transversal
photon spin conserving, form factors
\begin{equation}\labeleq{ffdvcs}
  \comptonT[\nu\mu]{}
  =\sum_{\lambda=\pm}
      \inbosonp[\lambda]{\nu}\outboson[\lambda]{\mu}
      \comptonT{\Tt}
    +\sum_{\lambda=\pm}
      \lambda\inbosonp[\lambda]{\nu}\outboson[\lambda]{\mu}
      \comptonTtilde{\Tt}
    +\ldots
\end{equation}
whose exact Lorentz structures will be defined when they are explicitly
needed in the next section.  For now it is sufficient to note that they
are equivalent to the usual ones in the generalized \Bjorken\ limit,
e.~g. the tensors $\tilde{t}_{\mu\nu}$ \{Eq.~(9) in
Ref.\cite{Belitsky:2000gz}\}.  These two form factors can be further
decomposed with respect to their Dirac structure
\begin{align}
  \comptonT{\Tt}
  &=\outFermion(P',S')\left[
      \comptonT{\Tt,1}(\xi,t,Q^2)
        \frac{\dirac{\qbar}}{2\dotprod{\Pbar}{\qbar}}
      +\comptonT{\Tt,2}(\xi,t,Q^2)
        \frac{i\sigma^{\mu\nu}\qbar_{\mu}\varDelta_{\nu}}{
          4M\dotprod{\Pbar}{\qbar}}
    \right]\inFermion(P,S)\mcomma\\
  \comptonTtilde{\Tt}
  &=\outFermion(P',S')\left[
      \comptonTtilde{\Tt,\Ta}(\xi,t,Q^2)
        \frac{\dirac{\qbar}}{2\dotprod{\Pbar}{\qbar}}
      +\comptonTtilde{\Tt,\Tp}(\xi,t,Q^2)
        \frac{\dotprod{\qbar}{\varDelta}}{4M\dotprod{\Pbar}{\qbar}}
    \right]\gamma_{5}\inFermion(P,S)
\end{align}
in analogy to the form factors of the electromagnetic current between
two nucleon states.

On the basis of the factorization theorems proven in
Refs.~\cite{Radyushkin:1997ki,Ji:1998xh,Collins:1998be} each of these
form factors can be expressed as convolutions of GPDs with
perturbatively calculable hard scattering coefficients such as
\begin{equation}\labeleq{fthm}
  \comptonT{\Tt}
  =\sum_{a=q,g}
    \intfthm{y}{-1}{+1}{\gpdf{a}(y,\xi,t,\mu)
      \sum_{b=q'}e_{b}^{2}\comptont{\Tt,ba}(\xi/y,Q^{2},\mu)}\mperiod
\end{equation}
Equivalent formulas can be obtained after the replacements
$\comptonT{\Tt}\to\comptonT{\Tt,1},\comptonT{\Tt,2}$ and
$\gpdf{a}\to\gpdH{a},\gpdE{a}$,
or $\comptonT{\Tt}\to\comptonTtilde{\Tt},
\comptonTtilde{\Tt,\Ta},\comptonTtilde{\Tt,\Tp}$,
$\gpdf{a}\to\gpdftilde{a},\gpdHtilde{a},\gpdEtilde{a}$, and
$\comptont{\Tt,ba}\to\comptonttilde{\Tt,ba}$.  The GPDs in
Eq.~\refeq{fthm} are defined in such a way that they have the closest
possible connection to the unpolarized and polarized DIS parton
distribution functions, that is
$\gpdH{q,g}(x,0,0,\mu)\equiv{}q,g(x,\mu)$ and
$\gpdHtilde{q,g}(x,0,0,\mu)\equiv{}\delta{}q,\delta{}g(x,\mu)$,
respectively.  The factorization scale $\mu$ is as usual assumed to be
equal to the renormalization scale with the common choice
$\mu^{2}=Q^{2}$.  The hard scattering coefficients describe the
scattering process $\gamma^{\virtual}a\to\gamma{}a$ of the intrinsic
parton $a$ with the coupling to the photon being mediated by the quark
$b$ and accordingly define the sums in the factorization theorems.

Crossing symmetry of the Compton amplitude allows to rewrite the hard
scattering coefficients in the following form
\begin{align}\labeleq{cfthm}
  \comptont{\Tt,ba}(z,Q^{2},\mu)
    &=\cfthm{\Tt,ba}(z,Q^{2},\mu)
      +\cfthm{\Tt,ba}(-z,Q^{2},\mu)\mcomma\notag\\
  \comptonttilde{\Tt,ba}(z,Q^{2},\mu)
    &=\ctildefthm{\Tt,ba}(z,Q^{2},\mu)
      -\ctildefthm{\Tt,ba}(-z,Q^{2},\mu)
\end{align}
where the definition of the coefficient function $\cfthm{\Tt,ba}$
($\ctildefthm{\Tt,ba}$) is not unique since odd (even) powers of $z$
drop out in the relevant combinations.  Eventually, the perturbative
expansion of the coefficient functions is written as
\begin{equation}\labeleq{cexpand}
  \cfthm{\Tt,ba}(z,Q^{2},\mu)=
    \cfthm[(0)]{\Tt,ba}(z,Q^{2},\mu)
    +\frac{\alphaqcd(\mu)}{2\pi}\cfthm[(1)]{\Tt,ba}(z,Q^{2},\mu)
    +\Landau(\alphaqcd^{2})\mperiod
\end{equation}

The \MSbar\ coefficient functions with exclusively massless parton lines
are available up to next-to-leading
order~\cite{Ji:1998xh,Belitsky:1999sg}.  For definiteness and since some
of them are needed for comparison in Sec.~\refsec{analytics} they are
explicitly given by
\begin{align}\labeleq{clfthmlo}
  \cfthm[(0)]{\Tt,ba}(z,Q^{2},\mu)
    &=\ctildefthm[(0)]{\Tt,ba}(z,Q^{2},\mu)
    =\kronecker_{ba}\frac{1}{z(1-i\ieps)-1}\mcomma\\
  \labeleq{clfthm}
  \cfthm[(1)]{\Tt,ba}(z,Q^{2},\mu)
    &=-\kronecker_{ba}\frac{\CF}{2}\left[
        \frac{9}{z-1}+\frac{3}{z+1}\log\frac{z-1}{2z}
        -\frac{1}{z-1}\log^{2}\frac{z-1}{2z}\right.\notag\\
    &\phantom{{}=-\kronecker_{ba}\frac{\CF}{2}}\left.{}
        -\frac{1}{z-1}\log\frac{Q^{2}}{\mu^{2}}
          \left(3+2\log\frac{z-1}{2z}\right)\right]\mcomma\notag\\
  \cfthm[(1)]{\Tt,bg}(z,Q^{2},\mu)
    &=\frac{\TF}{2(z+1)^{2}}\log\frac{z-1}{2z}\left[
        \frac{4}{z-1}+6-\log\frac{z-1}{2z}
        -2\log\frac{Q^{2}}{\mu^{2}}\right]\mcomma\notag\\
  \ctildefthm[(1)]{\Tt,ba}(z,Q^{2},\mu)
    &=\cfthm[(1)]{\Tt,ba}(z,Q^{2},\mu)
      +\kronecker_{ba}\frac{\CF}{z+1}\log\frac{z-1}{2z}\mcomma\notag\\
  \ctildefthm[(1)]{\Tt,bg}(z,Q^{2},\mu)
    &=\frac{\TF}{2(z+1)^{2}}\log\frac{z-1}{2z}\left[
        \frac{4}{z-1}-2+\log\frac{z-1}{2z}
        +2\log\frac{Q^{2}}{\mu^{2}}\right]\mperiod
\end{align}
None of the coefficient functions that contain internal massive quark
lines have been available so far and it is the aim of the next two
sections to obtain these in lowest order.

In the general form of the factorization theorem in Eq.~\refeq{fthm} the
inclusion of heavy quarks is simply done by appropriate specification of
the GPDs and the accompanying hard scattering coefficients.  In the
fixed order perturbation theory approach one has GPDs for the three
light flavors $u,d,s$ and for gluons.  In the hard scattering
coefficients the light flavors are considered massless whereas the
masses of the heavy quarks are kept.  Explicitly, in leading order one
has
\begin{align}\labeleq{fthmlo}
  \comptonT{\Tt}=\intfthm{y}{-1}{+1}{}
    &\left[
      \sum_{l=u,d,s}
        e_{l}^{2}\gpdf{l}(y,\xi,t,\mu)
        \comptont[(0)]{\Tt,ll}(\xi/y,Q^{2},\mu)\right.\notag\\
    &\left.
      +\sum_{h=c,b,t}
        e_{h}^{2}\gpdf{g}(y,\xi,t,\mu_{h})
        \comptont[(1)]{\Tt,hg}(\xi/y,Q^{2},\mu_{h})\right]
\end{align}
with an analogous formula for the helicity dependent form factor
$\comptonTtilde{\Tt}$.  The factorization scale of the fixed order part
can be chosen independently and is preferably
$\mu_{h}=2\mhq$~\cite{Gluck:1994dp}.  Apparently there seems to be a
mismatch in the order of the hard scattering coefficients as the lowest
order massive ones start at $\Landau(\alphaqcd)$ that can be resolved by
comparing the respective magnitudes which turn out to be partly of the
same size.  Additionally this is in line with the \massless\ parton
approach where heavy quarks contribute on the same level as the light
quarks.  In this latter case one considers GPDs for all quark flavors
where the intrinsic heavy quark GPDs are generated by the usual massless
evolution equations~\cite{Muller:1994fv,Ji:1997nm,Belitsky:1999hf} with
the boundary conditions, in analogy to Ref.~\cite{Collins:1986mp},
\begin{equation}\labeleq{mpbc}
  \gpdf{h}(x,\xi,t,\mu<\mhq)\equiv0\mlist
  \gpdftilde{h}(x,\xi,t,\mu<\mhq)\equiv0
\end{equation}
and the massless coefficient functions are used throughout.
Specifically in Eq.~\refeq{fthmlo} the first sum is extended to all
flavors and the second sum with the gluon GPD is absent.
%
%
\section{Photon-gluon scattering}\labelsec{pgs}
The lowest-order Feynman diagrams that contribute to photon-gluon
scattering are shown in Fig.~\reffig{pgs}.
\begin{figure}
  \renewcommand{\thesubfigure}{}
  \subfigure[(a)]{%
    \feynmangraph{30}{24}{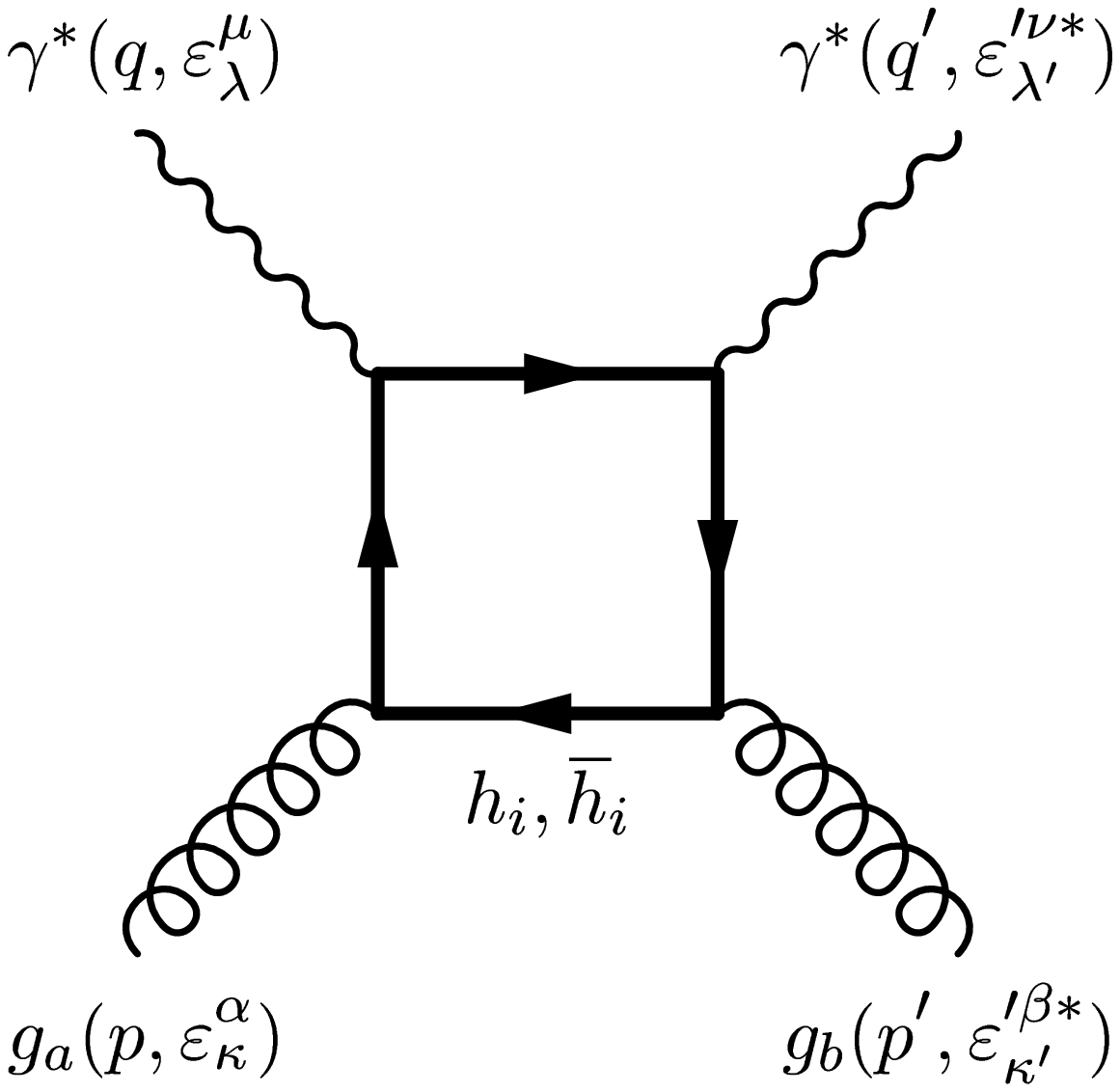}{
      \fmfleft{ig,ip}
        \fmflabel{\makebox[0pt]{\small%
          $\quad{}g_{a}(p,\inboson[\kappa]{\alpha})%
          \vphantom{\outbosonp[\kappa']{\beta}}$}}{ig}
        \fmflabel{\makebox[0pt]{\small%
          $\quad\gamma^{\virtual}(q,\inboson[\lambda]{\mu})%
          \vphantom{q'^{\vphantom{\mu}}_{\vphantom{\lambda'}}}$}}{ip}
      \fmfright{og,op}
        \fmflabel{\makebox[0pt]{\small%
          $g_{b}(p'^{\vphantom{\beta}}_{\vphantom{\kappa'}},%
          \outbosonp[\kappa']{\beta})\quad$}}{og}
        \fmflabel{\makebox[0pt]{\small%
          $\gamma^{\virtual}(q'^{\vphantom{\mu}}_{\vphantom{\lambda'}},%
          \outbosonp[\lambda']{\nu})\quad$}}{op}
      \fmfpen{thin}
        \fmf{phantom}{vig,ig}
        \fmf{phantom}{ip,vip}
        \fmf{phantom}{og,vog}
        \fmf{phantom}{vop,op}
        \fmf{fermion,tension=.707,width=thick}{vig,vip,vop,vog,vig}
      \fmffreeze
        \fmf{gluon}{vig,ig}
        \fmf{photon}{ip,vip}
        \fmf{gluon}{og,vog}
        \fmf{photon}{vop,op}
        \fmf{phantom,label.side=left,label=\small%
          $h_{i},,\anti{h}_{i}$}{vog,vig}
    }
  }
  \hspace{-10pt}
  \hspace{\fill}
  \subfigure[(b)]{%
    \feynmangraph{30}{24}{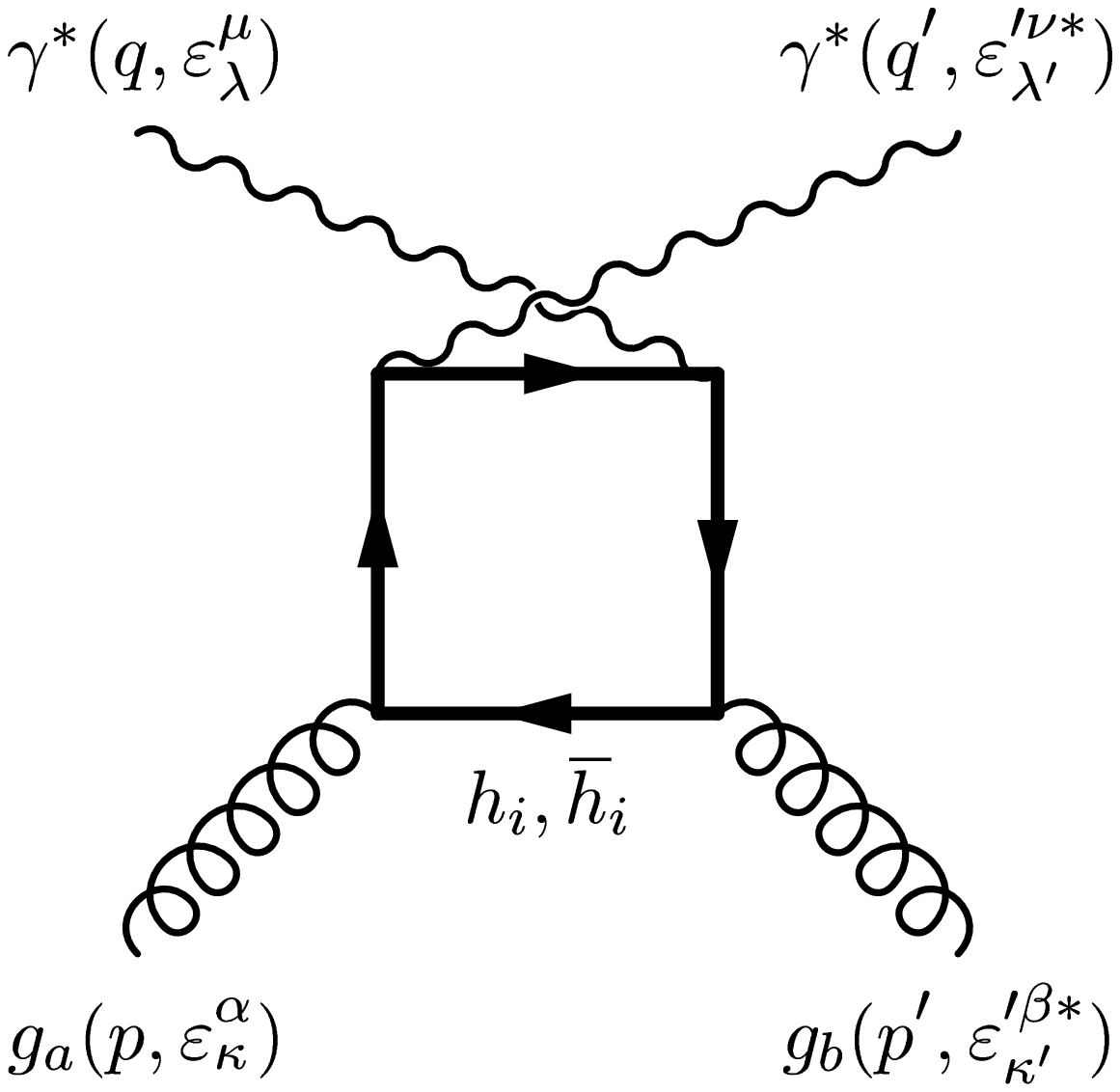}{
      \fmfleft{ig,ip}
        \fmflabel{\makebox[0pt]{\small%
          $\quad{}g_{a}(p,\inboson[\kappa]{\alpha})%
          \vphantom{\outbosonp[\kappa']{\beta}}$}}{ig}
        \fmflabel{\makebox[0pt]{\small%
          $\quad\gamma^{\virtual}(q,\inboson[\lambda]{\mu})%
          \vphantom{q'^{\vphantom{\mu}}_{\vphantom{\lambda'}}}$}}{ip}
      \fmfright{og,op}
        \fmflabel{\makebox[0pt]{\small%
          $g_{b}(p'^{\vphantom{\beta}}_{\vphantom{\kappa'}},%
          \outbosonp[\kappa']{\beta})\quad$}}{og}
        \fmflabel{\makebox[0pt]{\small%
          $\gamma^{\virtual}(q'^{\vphantom{\mu}}_{\vphantom{\lambda'}},%
          \outbosonp[\lambda']{\nu})\quad$}}{op}
       \fmfpen{thin}
        \fmf{phantom}{vig,ig}
        \fmf{phantom}{ip,vip}
        \fmf{phantom}{og,vog}
        \fmf{phantom}{vop,op}
        \fmf{fermion,tension=.707,width=thick}{vig,vip,vop,vog,vig}
      \fmffreeze
        \fmf{gluon}{vig,ig}
        \fmf{photon}{ip,vop}
        \fmf{gluon}{og,vog}
        \fmf{photon,rubout}{vip,op}
        \fmf{phantom,label.side=left,label=\small%
          $h_{i},,\anti{h}_{i}$}{vog,vig}
    }
  }
  \hspace{-10pt}
  \hspace{\fill}
  \subfigure[(c)]{%
    \feynmangraph{30}{24}{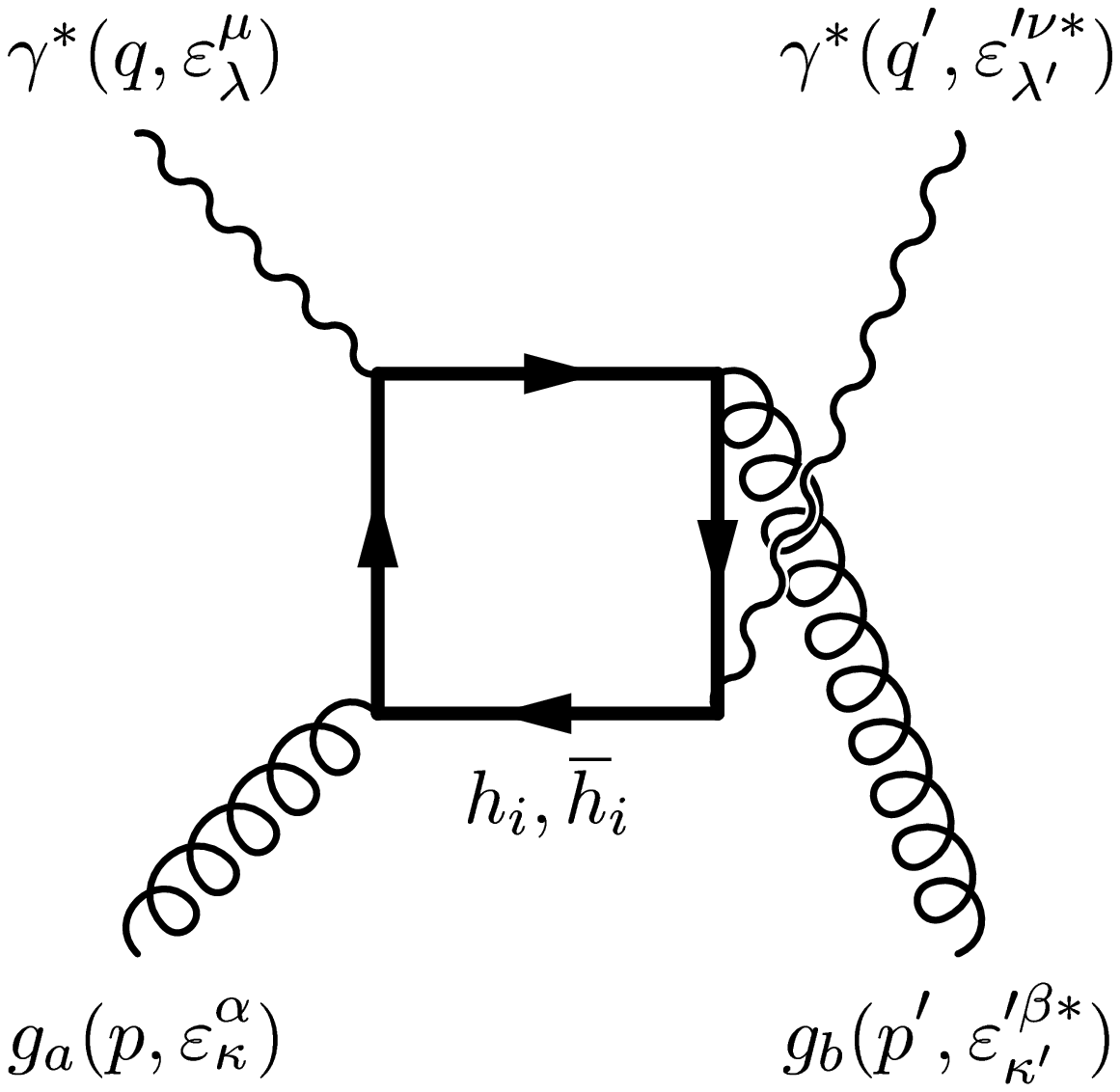}{
      \fmfleft{ig,ip}
        \fmflabel{\makebox[0pt]{\small%
          $\quad{}g_{a}(p,\inboson[\kappa]{\alpha})%
          \vphantom{\outbosonp[\kappa']{\beta}}$}}{ig}
        \fmflabel{\makebox[0pt]{\small%
          $\quad\gamma^{\virtual}(q,\inboson[\lambda]{\mu})%
          \vphantom{q'^{\vphantom{\mu}}_{\vphantom{\lambda'}}}$}}{ip}
      \fmfright{og,op}
        \fmflabel{\makebox[0pt]{\small%
          $g_{b}(p'^{\vphantom{\beta}}_{\vphantom{\kappa'}},%
          \outbosonp[\kappa']{\beta})\quad$}}{og}
        \fmflabel{\makebox[0pt]{\small%
          $\gamma^{\virtual}(q'^{\vphantom{\mu}}_{\vphantom{\lambda'}},%
          \outbosonp[\lambda']{\nu})\quad$}}{op}
       \fmfpen{thin}
        \fmf{phantom}{vig,ig}
        \fmf{phantom}{ip,vip}
        \fmf{phantom}{og,vog}
        \fmf{phantom}{vop,op}
        \fmf{fermion,tension=.707,width=thick}{vig,vip,vop,vog,vig}
      \fmffreeze
        \fmf{gluon}{vig,ig}
        \fmf{photon}{ip,vip}
        \fmf{gluon}{og,vop}
        \fmf{photon,rubout}{vog,op}
        \fmf{phantom,label.side=left,label=\small%
          $h_{i},,\anti{h}_{i}$}{vog,vig}
    }
  }
  \caption{Lowest order Feynman diagrams for photon-gluon scattering
    with color indices $a,b,i$.  The diagrams that have the heavy quark
    $h_{i}$ direction reversed are indicated by $\anti{h}_{i}$.}
  \labelfig{pgs}
\end{figure}
The diagrams with the quark direction reversed are not shown separately
but are indicated as antiquark loops.  Even though the reversion leads
to topologically different diagrams in configuration space, their
evaluation leads to the same result and is taken into account by an
overall factor of two in the following PGS amplitude in $D$ dimensions
\begin{align}\labeleq{pgs}
  \detachedlhs{
    \comptont[\beta\nu\alpha\mu]{ba}
    =ie_{h}^{2}g^{2}\kronecker_{ba}\TF\intloop[D]{k}{}}
    &\times2\trace
      \left[
        \frac{
          \gamma^{\alpha}(\dirac{k}+\mhq)
          \gamma^{\beta}(\dirac{k}+\diracp{p}+\mhq)
          \gamma^{\nu}(\dirac{k}+\diracp{p}+\diracp{q}+\mhq)
          \gamma^{\mu}(\dirac{k}+\dirac{p}+\mhq)
        }{
          \tiD_{k}(0,\mhq)\tiD_{k}(p',\mhq)
          \tiD_{k}(p'+q',\mhq)\tiD_{k}(p,\mhq)
        }
      \right.\notag\\
    &\phantom{\times\trace}\left.{}
        +\frac{
          \gamma^{\alpha}(\dirac{k}+\mhq)
          \gamma^{\beta}(\dirac{k}+\diracp{p}+\mhq)
          \gamma^{\mu}(\dirac{k}+\dirac{p}-\diracp{q}+\mhq)
          \gamma^{\nu}(\dirac{k}+\dirac{p}+\mhq)
        }{
          \tiD_{k}(0,\mhq)\tiD_{k}(p',\mhq)
          \tiD_{k}(p-q',\mhq)\tiD_{k}(p,\mhq)
        }
      \right.\notag\\
    &\phantom{\times\trace}\left.{}
        +\frac{
          \gamma^{\alpha}(\dirac{k}+\mhq)
          \gamma^{\nu}(\dirac{k}+\diracp{q}+\mhq)
          \gamma^{\beta}(\dirac{k}+\diracp{p}+\diracp{q}+\mhq)
          \gamma^{\mu}(\dirac{k}+\dirac{p}+\mhq)
        }{
          \tiD_{k}(0,\mhq)\tiD_{k}(q',\mhq)
          \tiD_{k}(p'+q',\mhq)\tiD_{k}(p,\mhq)
        }
      \right]
\end{align}
with $\tiD_{k}(p,\mhq)\defeq(k+p)^{2}-\mhq^{2}+i\ieps$.  The calculation
of this amplitude was carried out with the help of
\MATHEMATICA~\cite{Mathematica:4.1} that provided an adequate handling
of the excessive algebra and the \TRACER~\cite{Jamin:1993dp} package
that was used to evaluate the trace and to contract Lorentz indices.
The individual steps that lead to the analytical results are detailed
below.

The integral over the loop momentum can be systematically reduced to
scalar integrals
\begin{equation}
  \intn[D]{k}{}\left[
      \tiD_{k}(0,m_{0})\tiD_{k}(p_{1},m_{1})\cdots\tiD_{k}(p_{N},m_{N})
    \right]^{-1}
\end{equation}
by a Passarino-Veltman decomposition~\cite{Passarino:1979jh} which is
described in a form more suitable for implementation in computer algebra
systems in Ref.~\cite{Denner:1993kt}.  The whole PGS amplitude in
Eq.~\refeq{pgs} is finite whereas the separate scalar integrals can be
divergent.  Therefore the general $D$ dimensions have to be kept to
regularize any divergences during intermediate steps.  Explicit formulas
for the scalar integrals are given in
Refs.~\cite{'tHooft:1979xw,Denner:1991qq,Denner:1993kt} in which they
are expressed in terms of complex (di)logarithms whose arguments are
generically built up by
\begin{equation}
  \tir_{s}\defeq\frac{\tibeta_{s}-1}{\tibeta_{s}+1}
  \quad\text{with}\quad
  \tibeta_{s}\defeq\sqrt{1-\frac{4\mhq^{2}}{s+i\ieps}}
\end{equation}
if all masses are equal as it is the case for the lowest order PGS
amplitude.

The immediate application of the Passarino-Veltman decomposition to
Eq.~\refeq{pgs} is in principle possible but not practical.  The amount
of work is considerably reduced if one gets rid of the four free Lorentz
indices first.  Inspired by Refs.~\cite{Budnev:1974de,Schienbein:2002wj}
covariant tensors with an obvious physical meaning are constructed that
appear as independent Lorentz structures in the PGS amplitude and
equally act as projectors onto them.  The three covariant four-vectors
\begin{gather}
  \rpveps_{2}^{\smash{\mu}}
  \defeq\frac{
      \eta'^{\mu}-\dotprod{\eta'}{\eta}\eta^{\mu}
      -\dotprod{\eta'}{\hat{q}}\hat{q}^{\mu}
    }{
      \sqrt{\pdotprod{\eta'}{\eta}^{2}-\pdotprod{\eta'}{\hat{q}}^{2}
        -\eta'^{2}}
    }\mlist
  \rpveps_{1}^{\smash{\mu}}
  \defeq-\lceps_{\kappa\lambda\phantom{\mu}\nu}^{
        \phantom{\kappa\lambda}\smash{\mu}}
      \eta^{\kappa}\hat{q}^{\lambda}\rpveps_{2}^{\nu}\mlist
  \rpveps_{0}^{\smash{\mu}}
  \defeq\frac{
      \dotprod{\eta}{q}q^{\mu}-q^{2}\eta^{\mu}
    }{
      \sqrt{-q^{2}}\sqrt{(\dotprod{\eta}{q})^{2}-q^{2}}
    }\\
  \text{with}\quad
  \hat{q}^{\mu}
  \defeq\frac{
      q^{\mu}-\dotprod{q}{\eta}\eta^{\mu}
    }{
      \sqrt{\pdotprod{q}{\eta}^{2}-q^{2}}
    }\mlist\eta^{2}=1\mlist\notag
\end{gather}
$\eta'$ defining the direction of $\rpveps_{2}$, and definite parity
serve as a basis.  One easily proves that they reduce to the usual
linear polarization vectors of the incoming photon in the frame
$\vec{\eta}=\nullvec$.  Now the sought-for tensors are simple products
of these polarization vectors and analogous ones for $\alpha,\beta,\mu$
with the sensible choice $\eta\sim\pbar+\qbar=p+q$ and $\eta'=\qbar$.
Note that the totally antisymmetric epsilon tensor always appears in
pairs as the PGS amplitude is parity-even.  These products are easily
generalized to $D$ dimensions if they are expressed by metric tensors.
Physically, this means a sum over polarization vectors that are
perpendicular to the scattering plane.  In principle this sum has to be
normalized by a factor $1/(D-3)$ which is, however, not necessary for
the finite PGS amplitude in Eq.~\refeq{pgs}.

The complex circular polarization vectors used in the expansion of the
Compton amplitude in Eq.~\refeq{ffdvcs} are given by
\begin{equation}
  \cpveps_{\pm}^{\smash{\mu}}
  \defeq\mp\frac{1}{\sqrt{2}}
    \left(
      \rpveps_{1}^{\smash{\mu}}
      \pm{}i\rpveps_{2}^{\smash{\mu}}
    \right)\mlist
  \cpveps_{0}^{\smash{\mu}}
  \defeq{}i\rpveps_{0}^{\smash{\mu}}\mperiod
\end{equation}
From a parton model point of view it looks advantageous to use
$\eta\sim\pbar$ because in this case these polarization vectors are
invariant under the replacement $\pbar\to{}y\pbar$.  Anyway, both
choices are equivalent on the leading twist level and lead to the same
hard scattering coefficients that are presented in the next section.

In order to extract the leading twist contribution, one has to consider
the limit $t\to0$ where some care has to be taken since factors up to
$t^{2}$ that appear in denominators which originate from the
normalization of the polarization vectors have to be canceled
thoroughly.  In this limit the dilogarithms drop out completely.
%
%
\section{Analytical results}\labelsec{analytics}
At the leading twist-two level the general Compton amplitude with
virtual photons in the initial and final state receives contributions of
gluons inside the proton to the following form factors
\begin{align}
  \comptonT[\nu\mu]{}
  ={}&\sum_{\lambda=\pm}
      \inbosonp[\lambda]{\nu}\outboson[\lambda]{\mu}
      \comptonT{\Tt}
    +\sum_{\lambda=\pm}
      \lambda\inbosonp[\lambda]{\nu}\outboson[\lambda]{\mu}
      \comptonTtilde{\Tt}
    +\inbosonp[0]{\nu}\outboson[0]{\mu}\comptonT{\Tl}\notag\\
  &+\sum_{\lambda=\pm}
      \inbosonp[-\lambda]{\nu}\outboson[\lambda]{\mu}
      \comptonT{\Ttflip}
    +\sum_{\lambda=\pm}
      \lambda\inbosonp[-\lambda]{\nu}\outboson[\lambda]{\mu}
      \comptonTtilde{\Ttflip}
    +\ldots
\end{align}
where the three additional terms in comparison to Eq.~\refeq{ffdvcs} are
related to longitudinally polarized photons ($\Tl$) and to photon
helicity flip ($\Ttflip$).  Due to the real photon in the final state
$\comptonT{\Tl}$ does not appear in the DVCS amplitude.  From the
calculation of PGS that has been sketched in the last section it is
possible to obtain the lowest-order massive hard scattering coefficients
appearing in factorization theorems for these form factors like in
Eq.~\refeq{fthm}.  Now it will be indicated how they are extracted from
the amplitude in Eq.~\refeq{pgs}.

Generally, the squared heavy quark charge already appears
in the factorization formula, the QCD coupling constant is
extracted in accordance to Eq.~\refeq{cexpand}, and one gets an
additional factor $1/[2(1-\xi^{2})]$ from the normalization of the gluon
GPDs~\cite{Collins:1997fb,Ji:1998xh}.  The massive hard scattering
coefficients relevant for the numerical analysis in the next section are
contained in the gluon polarization vector sums
$\sum_{\lambda=\pm}\inbosonp[\lambda]{\beta}\outboson[\lambda]{\alpha}$
for $\comptonT{\Tt}$, resp.\ $\sum_{\lambda=\pm}
\lambda\inbosonp[\lambda]{\beta}\outboson[\lambda]{\alpha}$ for
$\comptonTtilde{\Tt}$ and read
\begin{align}\labeleq{tthg}
  \comptont[(1)]{\Tt,hg}
    &=-\frac{\TF}{2(1-\xi^{2})^{2}}\left\{
      \left[x^{2}-\xi^{2}+(1-x)^{2}+4\etabarh{}x(1-x)
          -8\etabarh^{2}x^{2}\right]
        \log^{2}\tir_{\shat}\right.\notag\\
    &\phantom{{}=-\frac{\TF}{2(1-\xi^{2})^{2}}}
      \vphantom{\frac{\TF}{(1-\xi^{2})^{2}}}\left.{}
      +2\left[(1-2x)^{2}-\xi^{2}+4\etabarh{}x(1-x)\right]
        \tibeta_{\shat}\log\tir_{\shat}\right.\notag\\
    &\phantom{{}=-\frac{\TF}{2(1-\xi^{2})^{2}}}
      \vphantom{\frac{\TF}{(1-\xi^{2})^{2}}}\left.{}
      -\left[\left(\vphantom{\xi^{2}}x/\xi+1-2\etabarh{}x/\xi\right)
          \left(1+2x\xi-\xi^{2}\right)
          -4\etabarh{}x\xi-8\etabarh^{2}x^{2}\right]
        \log^{2}\tir_{q^{2}}\right.\notag\\
    &\phantom{{}=-\frac{\TF}{2(1-\xi^{2})^{2}}}
      \vphantom{\frac{\TF}{(1-\xi^{2})^{2}}}\left.{}
      -2\left(\vphantom{\xi^{2}}x/\xi+1\right)
        \left(1+4x\xi-\xi^{2}-4\etabarh{}x\xi\right)
        \tibeta_{q^{2}}\log\tir_{q^{2}}
    +(x\to{-x})\right\}\notag\\
  \comptonttilde[(1)]{\Tt,hg}
    &=\frac{\TF}{2(1-\xi^{2})^{2}}\left[
      \left(1-2x+\xi^{2}\right)\log^{2}\tir_{\shat}
      +2\left(3-4x+\xi^{2}\right)
        \tibeta_{\shat}\log\tir_{\shat}\right.\notag\\
    &\phantom{{}=\frac{\TF}{2(1-\xi^{2})^{2}}}
      \vphantom{\frac{\TF}{(1-\xi^{2})^{2}}}\left.{}
      +2\left(\vphantom{\tibeta_{q^{2}}}x+\xi\right)
        \left(\log^{2}\tir_{q^{2}}
          +4\tibeta_{q^{2}}\log\tir_{q^{2}}\right)
      -(x\to{-x})\right]
\end{align}
with $\etabarh\defeq{}\mhq^{2}/\Qbar^{2}$.  The corresponding results
for massless quarks~\cite{Ji:1998xh,Belitsky:1999sg} are obtained for
$\etabarh\to0$ with appropriate subtraction of the logarithmic
divergences $\sim\log(\mhq/\mu)$.  It should be remarked that the limits
have to be taken in this order, namely, first $t\to0$ and then
$\etabarh\to0$, to reproduce the massless results.

Furthermore, the contraction of the gluonic Lorentz indices analogously
to $\comptonT{\Tt}$ gives the contribution of longitudinally polarized
photons
\begin{align}\labeleq{tlhg}
  \comptont[(1)]{\Tl,hg}
    =\frac{2\TF\sqrt{\abs{1-\xi^{2}/x^{2}}}}{(1-\xi^{2})^{2}}
    &\left[
      2\etabarh{}x^{2}
        \left(\log^{2}\tir_{\shat}-\log^{2}\tir_{q^{2}}\right)
      +2x(1-x)\tibeta_{\shat}\log\tir_{\shat}\right.\notag\\
    &\vphantom{\frac{\TF}{(1-\xi^{2})^{2}}}\left.{}
      +\left(x/\xi+2x^{2}+x\xi\right)\tibeta_{q^{2}}\log\tir_{q^{2}}
      +(x\to{-x})\right]\mperiod
\end{align}
The \massless\ limit can be achieved in the same way as mentioned above
but will not be explicitly stated since it is not needed at present.

As a byproduct of the calculation of the PGS amplitude with the method
presented in the previous section one also gets massive expressions for
the hard scattering coefficients in factorization theorems with
generalized helicity-flip, or tensor gluon distributions, respectively,
whose exact definitions and applications are given in
Refs.~\cite{Hoodbhoy:1998vm,Belitsky:2000jk,Diehl:2001pm}.  They have to
be projected from the gluon polarization vector sums
$\sum_{\lambda=\pm}\inbosonp[-\lambda]{\beta}\outboson[\lambda]{\alpha}$
for $\comptonT{\Ttflip}$ and $\sum_{\lambda=\pm}
\lambda\inbosonp[-\lambda]{\beta}\outboson[\lambda]{\alpha}$ for
$\comptonTtilde{\Ttflip}$ and are given for completeness
\begin{align}
  \comptont[(1)]{\Ttflip,hg}
    =\comptonttilde[(1)]{\Ttflip,hg}=\frac{2\TF}{(1-\xi^{2})^{2}}
    &\left[
      2\etabarh(1+\etabarh)x^{2}
        \left(\log^{2}\tir_{\shat}-\log^{2}\tir_{q^{2}}\right)
      -2\etabarh{}x(1+\xi)\tibeta_{\shat}\log\tir_{\shat}\right.\notag\\
    &\vphantom{\frac{\TF}{(1-\xi^{2})^{2}}}\left.{}
      -\left(\vphantom{\tibeta_{q^{2}}}x+\xi\right)
        \left(\vphantom{\tibeta_{q^{2}}}x-\xi-2\etabarh{}x\right)
        \left(\tibeta_{\shat}\log\tir_{\shat}
          -\tibeta_{q^{2}}\log\tir_{q^{2}}\right)
      +(x\to{-x})\right]\notag\\
    {}+\frac{2\TF}{1-\xi^{2}}&\mperiod
\end{align}
The limit $\etabarh\to0$ is finite as expected because the generalized
tensor gluon distributions do not mix with any quark distributions and
reproduces the \massless\ results of
Refs.~\cite{Hoodbhoy:1998vm,Belitsky:2000jk}.

Two further limits provide additional checks for the mass dependencies
in Eq.~\refeq{tthg}.  On the one hand the contributions of heavy quarks
do vanish for infinitely large masses in accordance with the decoupling
theorem~\cite{Appelquist:1975tg}.  On the other hand the imaginary parts
of Eqs.~\refeq{tthg} and \refeq{tlhg} in the forward scattering limit
are given by
\begin{align}
  &\frac{1}{\pi}\imagpart{}\comptont[(1)]{\Tt,hg}
    \xrightarrow{\xi\to{}0}-\TF\MOtheta(1/x-1-4\etah)\notag\\
    &\qquad\vphantom{\frac{1}{\pi}}\times\left\{
      \left[x^{2}+(1-x)^{2}+4\etah{}x(1-x)-8\etah^{2}x^{2}\right]
        \log(-\tir_{\shat})
      +\left[(1-2x)^{2}+4\etah{}x(1-x)\right]\tibeta_{\shat}\right\}
  \notag\\
  &\frac{1}{\pi}\imagpart{}\comptont[(1)]{\Tl,hg}
    \xrightarrow{\xi\to{}0}4\TF\MOtheta(1/x-1-4\etah)\left[
      2\etah{}x^{2}\log(-\tir_{\shat})+x(1-x)\tibeta_{\shat}\right]
  \notag\\
  &\frac{1}{\pi}\imagpart{}\comptonttilde[(1)]{\Tt,hg}
    \xrightarrow{\xi\to{}0}\TF\MOtheta(1/x-1-4\etah)
      \left[(1-2x)\log(-\tir_{\shat})+(3-4x)\tibeta_{\shat}\right]
\end{align}
with $\etah\defeq{}\mhq^{2}/Q^{2}$. They coincide with the corresponding
quantities of the DIS structure functions $(F_{2}-F_{L})/x$,
$F_{L}/x$~\cite{Witten:1976bh}, and $2g_{1}$~\cite{Gluck:1991in} in
view of the factorization theorem Eq.~\refeq{fthm} as demanded by the
optical theorem.

Finally the relevant results for the numerical analysis in the following
section are the massive DVCS coefficient functions as defined by
Eq.~\refeq{cfthm}
\begin{align}\labeleq{chfthm}
  \cfthm[(1)]{\Tt,hg}(z,Q^{2},\mu)=\frac{\TF}{2(z+1)^{2}}
    &\left\{
      \frac{2-z(6-16\eta)}{z-1}\left(\tibeta_{\shat}\log\tir_{\shat}
        -\tibeta_{q^{2}}\log\tir_{q^{2}}\right)\right.\notag\\
    &\vphantom{\frac{\TF}{(1-\xi^{2})^{2}}}\left.{}
      -\left[1+8\eta{}z\frac{1-(1+4\eta)z}{(z-1)^{2}}\right]
        \log^{2}\tir_{\shat}\right.\notag\\
    &\vphantom{\frac{\TF}{(1-\xi^{2})^{2}}}\left.{}
      +\left[1-2\eta\frac{1+(3+8\eta)z^{2}}{(z-1)^{2}}\right]
        \log^{2}\tir_{q^{2}}\right\}\notag\\
  \ctildefthm[(1)]{\Tt,hg}(z,Q^{2},\mu)=\frac{\TF}{2(z+1)^{2}}
    &\left[
      \frac{2z-6}{z-1}\left(\tibeta_{\shat}\log\tir_{\shat}
        -\tibeta_{q^{2}}\log\tir_{q^{2}}\right)
      +\log^{2}\tir_{\shat}-\log^{2}\tir_{q^{2}}\right]
\end{align}
that are achieved by enforcing the DVCS kinematics $x\to\xi$.  In the
combinations of Eq.~\refeq{cfthm} they are finite for $z\to1$ and the
imaginary parts stem exclusively from the complex logarithms.  The
arbitrariness in the representations of DVCS coefficient functions is
exploited so that Eq.~\refeq{chfthm} reduces to its equivalent in
Eq.~\refeq{clfthm} for $\eta\to0$ apart from the logarithmic divergences
$\sim\log(\mhq/\mu)$.
%
%
\section{Numerical results}\labelsec{numerics}
A suitable model for the GPDs to employ in a first numerical analysis of
heavy quark effects is given by the moment-diagonal model defined in
Ref.~\cite{Noritzsch:2000pr}.  One can expect that it leads to the same
difficulties to describe experimental data, which is explained in detail
in Ref.~\cite{Freund:2002qf}, as factorized double distribution based
models for GPDs~\cite{Radyushkin:1998bz}.  Nevertheless moment diagonal
models have important advantages.  They are scale independently related
to the conventional forward parton distribution functions (PDFs) so that
no additional GPD input scale uncertainty is introduced and their
dependence on the specific forward PDF set is reduced.  Therefore, the
following results should be unique consequences of heavy
quark contributions.  Apart from that, the evaluation of the form
factors of the DVCS amplitude can be performed in a numerically very
stable way without any principal value
integrals~\cite{Noritzsch:2000pr}. The utilized unpolarized and
polarized forward PDFs are GRV(98)~\cite{Gluck:1998xa} and the standard
scenario of GRSV(00)~\cite{Gluck:2000dy}, respectively, with three fixed
light flavors.  The charm and bottom quark masses are $m_{c}=1.4\,\GeV$
and $m_{b}=4.5\,\GeV$.  In accordance to \refeq{mpbc} and
Ref.~\cite{Collins:1986mp}, intrinsic \massless\ (off-)forward heavy
quark distribution functions were generated from the same inputs.  It
should be noted that all the following fixed-order (FO) and \massless\
parton (MP) results for the three light quarks $u,d,s$ agree within
$1.5\%$.

In Fig.~\reffig{tt3}
\begin{figure}
  \includegraphics[angle=-90,width=49\unitlength]{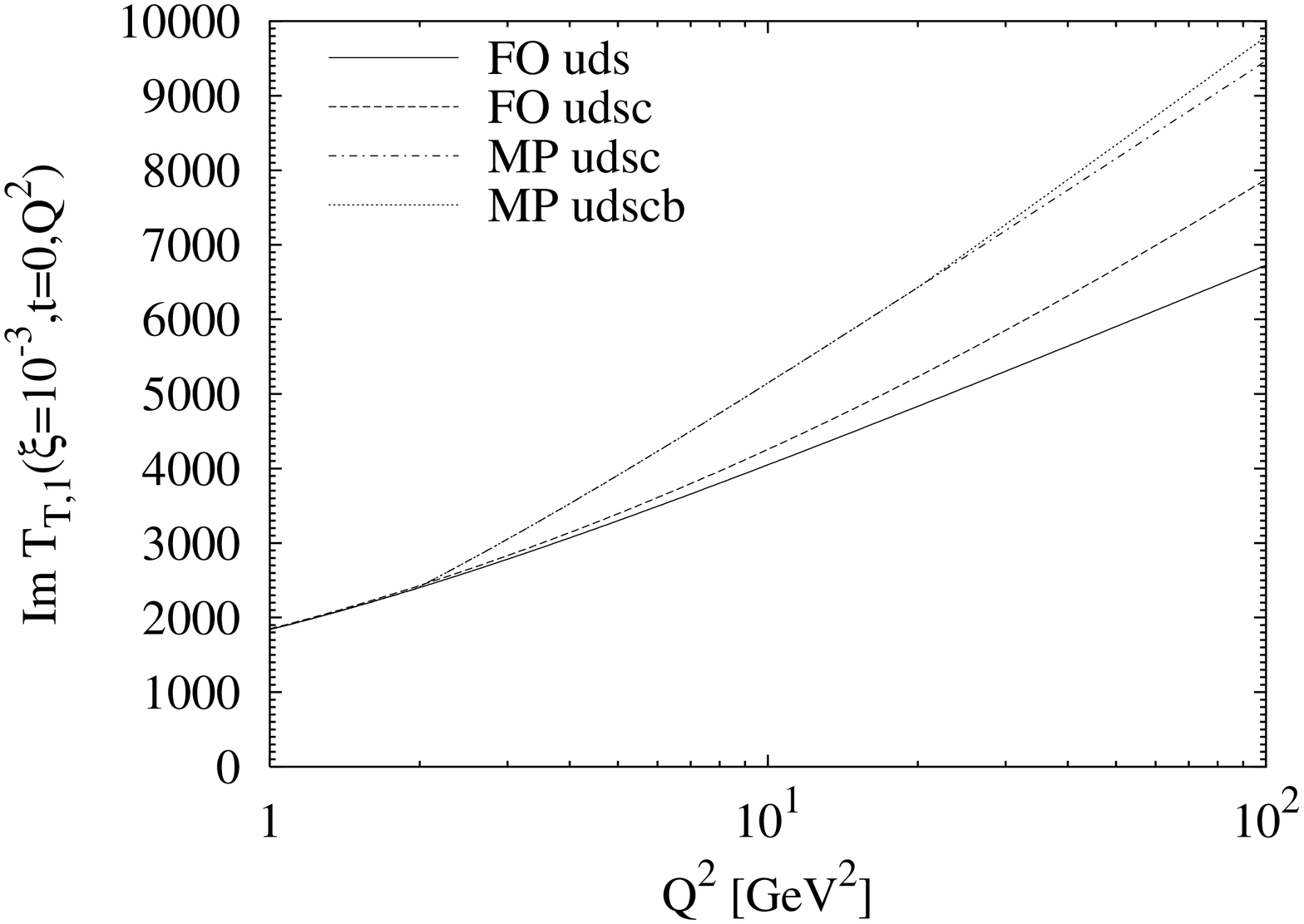}
  \hspace{\fill}
  \includegraphics[angle=-90,width=49\unitlength]{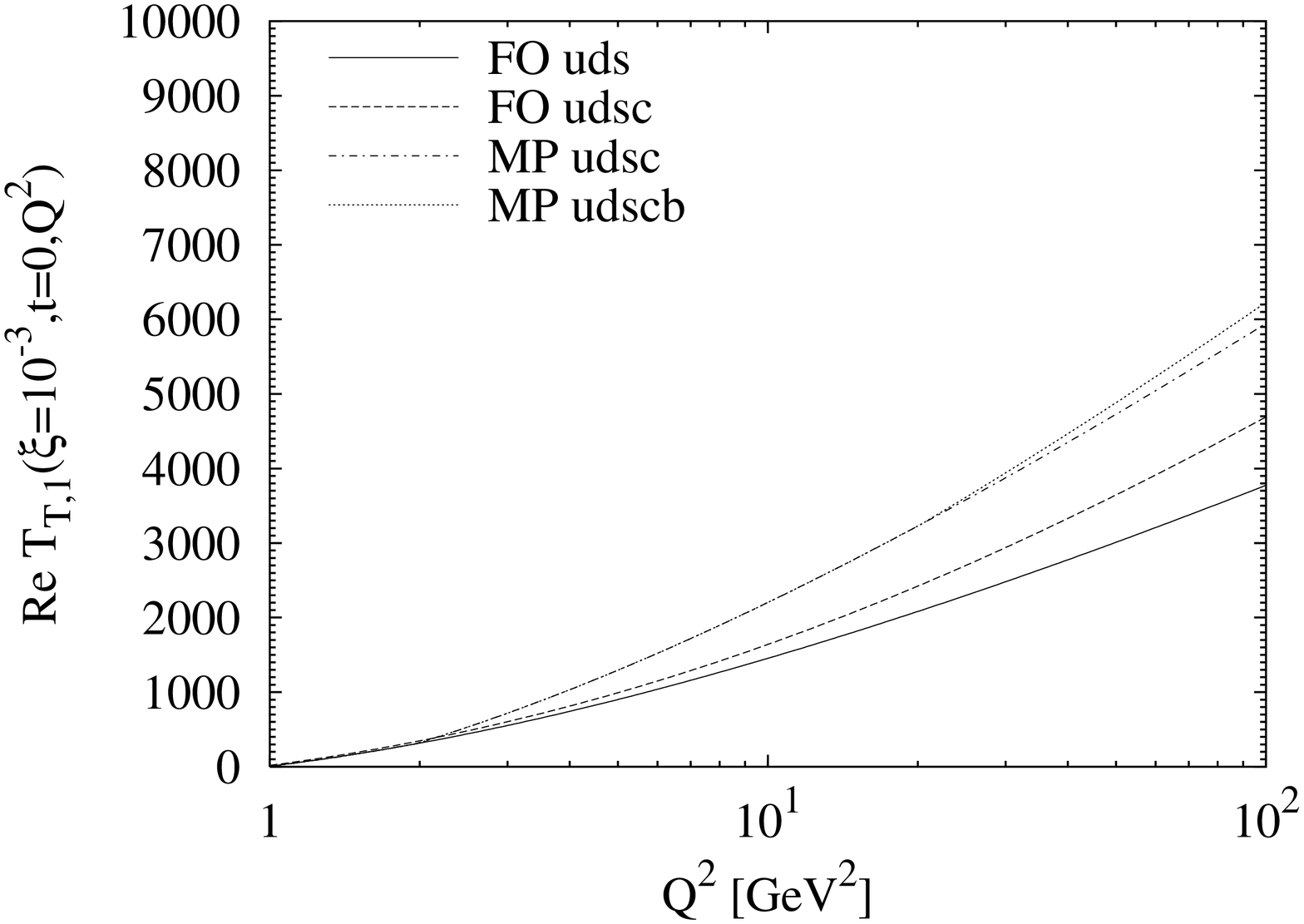}
  \caption{The imaginary and real part of the form factor
    $\comptonT{\Tt,1}$, related to $\gpdH{}$ in the appropriately
    substituted Eq.~\refeq{fthm}, of the DVCS amplitude as a function of
    $Q^{2}$ at fixed $\xi=10^{-3}$ and $t=0$ for three light flavors
    (solid line), with the fixed order charm contribution (dashed line),
    with the \massless\ charm (dashed-dotted line) and additionally
    bottom (dotted line) contribution.}
  \labelfig{tt3}
\end{figure}
the results for the form factor $\comptonT{\Tt,1}(\xi,t,Q^{2})$ of the
DVCS amplitude, which is the one that depends on $\gpdH{}(x,\xi,t,\mu)$,
are presented as a function of $Q^{2}$ for vanishing $t$ and fixed
$\xi=10^{-3}$ which is representative for the kinematic region of DVCS
measurements by H1~\cite{Adloff:2001cn} and by
ZEUS~\cite{Chekanov:2003ya} at DESY-HERA.  The contributions of the
intrinsic \massless\ heavy quark GPDs show as expected a different
threshold behavior compared to the fixed order results.  The former
start rather abrupt at $Q^{2}=\mhq^{2}$ in the imaginary and real part
whereas the latter set in smoothly at $\shat>4\mhq^{2}$ in the
imaginary part and contributes to the real part at any scale, though not
sizeably for $Q^{2}\ll\mhq^{2}$.  Between the threshold region and
the highest $Q^{2}=100\,\GeV^{2}$ displayed in Fig.~\reffig{tt3} with
the \massless\ charm quark contribution in \texttt{MP\,udsc} lies
significantly above the fixed order result \texttt{FO\,udsc}.  The small
\massless\ bottom quark contribution justifies to neglect all further
bottom results from now on, as it has been already done for the nearly
vanishing fixed order bottom results in Fig.~\reffig{tt3}.

The corresponding results with fixed $\xi=10^{-1}$ relevant for the
kinematic region of HERMES~\cite{Airapetian:2001yk} and
CLAS~\cite{Stepanyan:2001sm} are shown in Fig.~\reffig{tt1}.
\begin{figure}
  \includegraphics[angle=-90,width=49\unitlength]{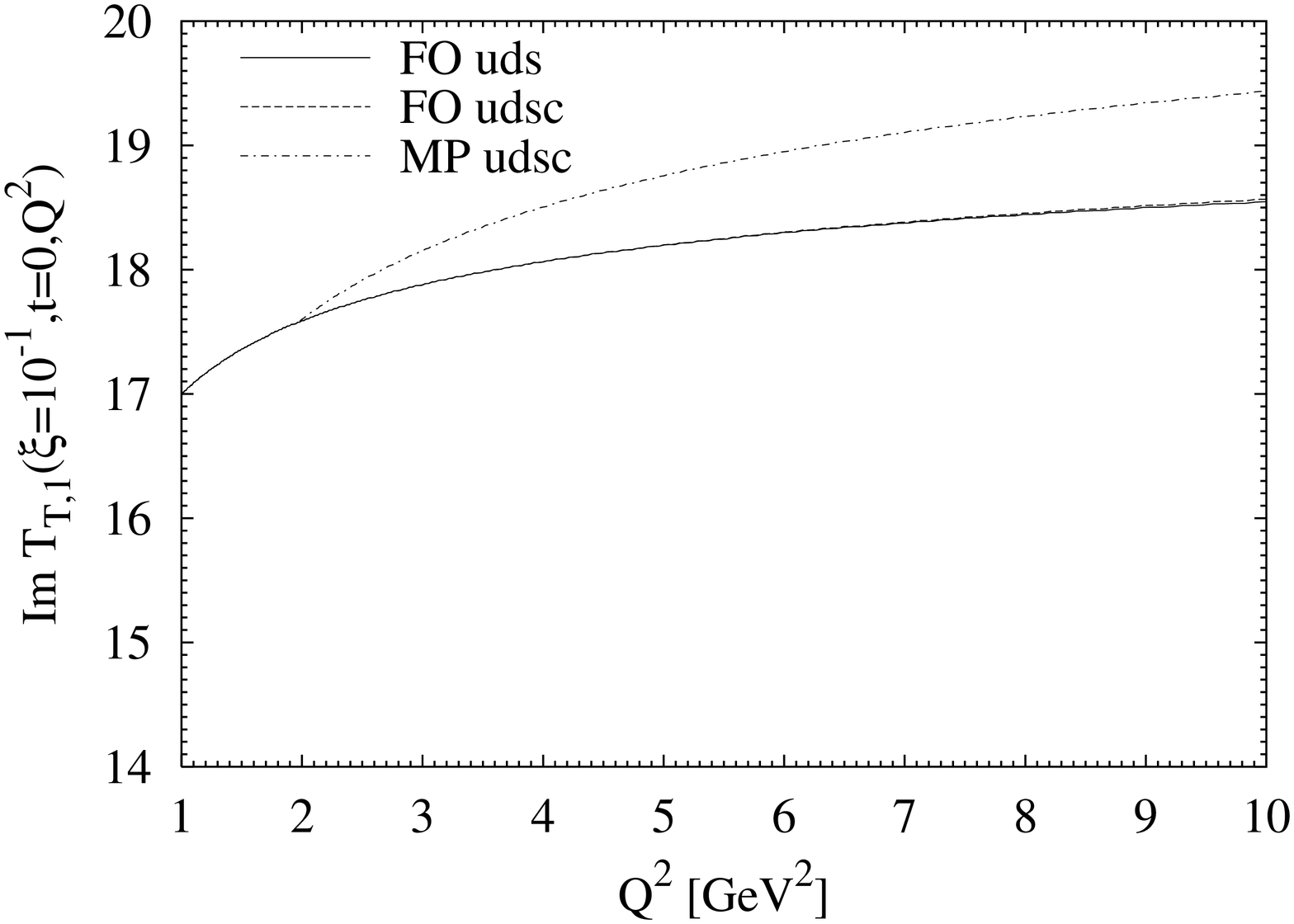}
  \hspace{\fill}
  \includegraphics[angle=-90,width=49\unitlength]{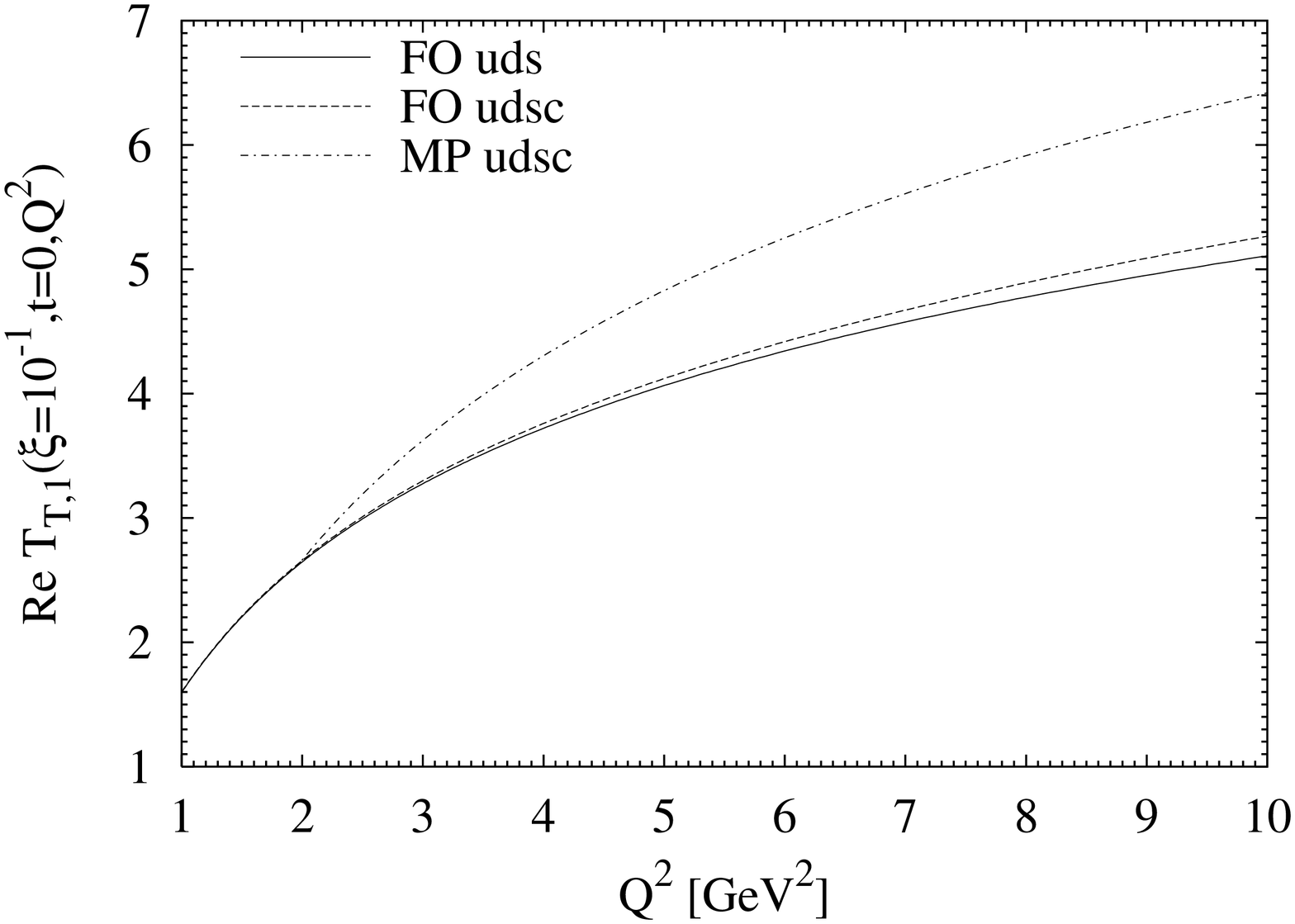}

  \includegraphics[angle=-90,width=49\unitlength]{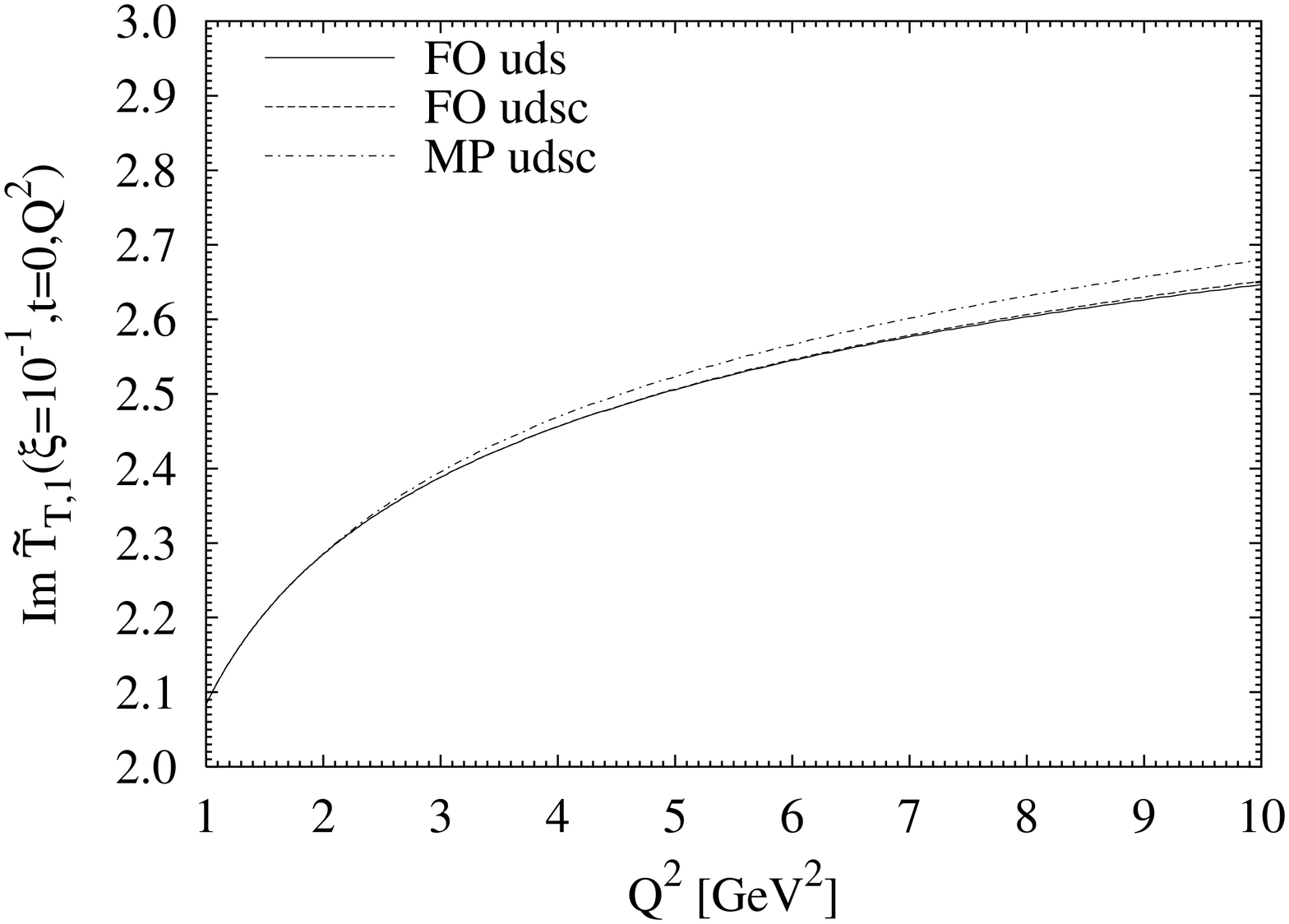}
  \hspace{\fill}
  \includegraphics[angle=-90,width=49\unitlength]{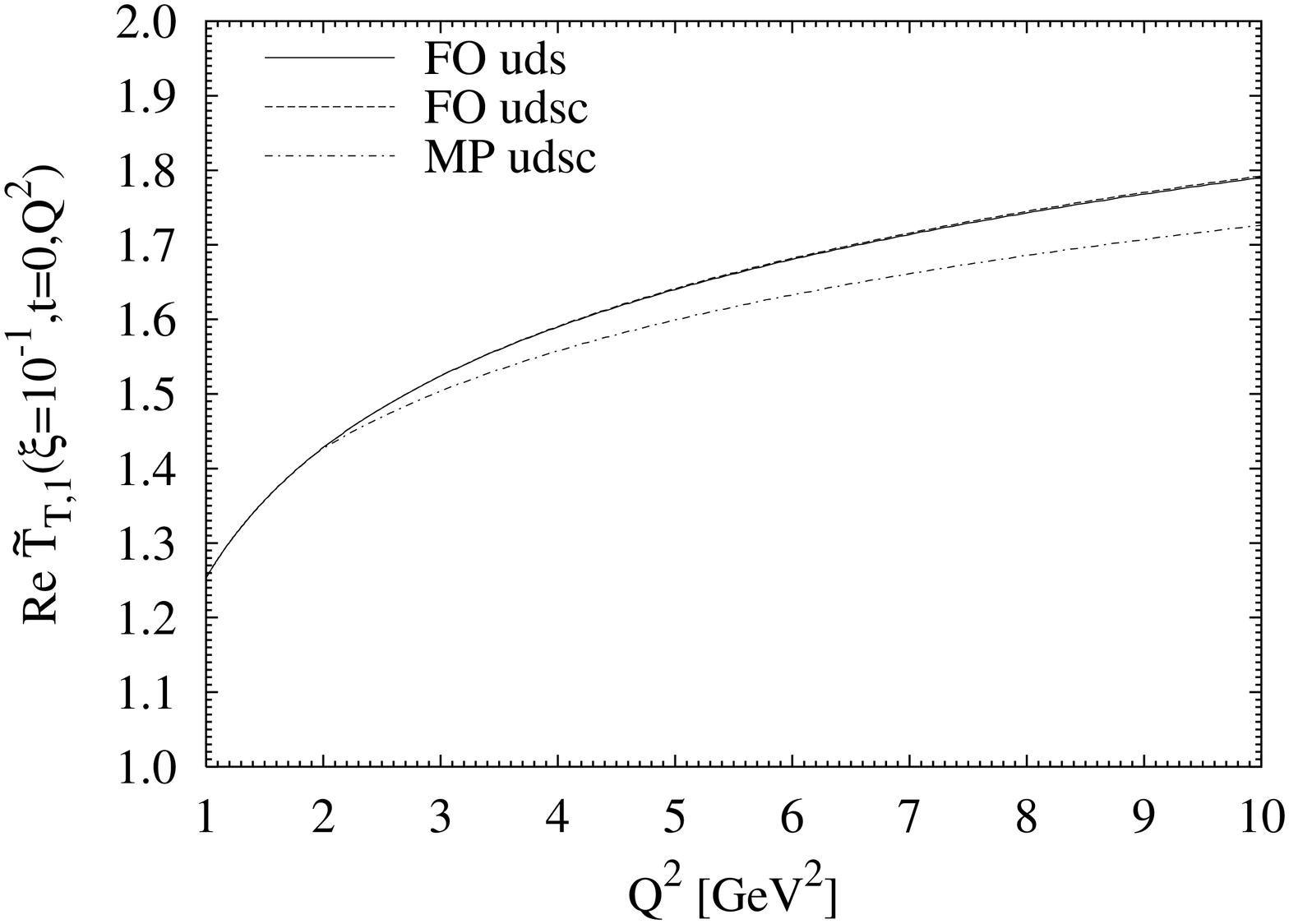}
  \caption{The imaginary and real parts of the form factors
    $\comptonT{\Tt,1}$ and $\comptonTtilde{\Tt,\Ta}$, related to
    $\gpdH{}$ and $\gpdHtilde{}$, respectively, of the DVCS amplitude as
    a function of $Q^{2}$ at fixed $\xi=10^{-1}$ and $t=0$ for three
    light flavors (solid line), with the fixed order charm contribution
    (dashed line), and with the \massless\ charm (dashed-dotted line)
    contribution.}
  \labelfig{tt1}
\end{figure}
In this case the helicity dependent form factor
$\comptonTtilde{\Tt,\Ta}(\xi,t,Q^{2})$, which is the one that depends on
$\gpdHtilde{}(x,\xi,t,\mu)$ in the appropriately substituted
Eq.~\refeq{fthm}, is also presented.  At even smaller values of $\xi$,
e.~g.\ $\xi=10^{-3}$ in Fig.~\reffig{tt3}, $\comptonTtilde{\Tt,\Ta}$
becomes entirely negligible compared to $\comptonT{\Tt,1}$.  The
\massless\ parton prescription still leads to considerable charm
contributions whereas the fixed order charm results are marginal.
Generally the corrections to DVCS observables due to heavy quarks get
larger for increasing $Q^{2}$ and in particular for decreasing $\xi$.
Therefore phenomenologically relevant effects have to be expected mainly
for the DVCS cross section measurements at H1 and ZEUS.

The DVCS cross section was defined by using the equivalent photon
approximation in Ref.~\cite{Adloff:2001cn}.  An approximate formula is
given by~\cite{Freund:2001hd}
\begin{equation}
  \sigma_{\text{DVCS}}
    \approx\frac{\pi\alpha^{2}\xBjorken^{2}}{Q^{4}b}
      \lrabs{\comptonT{\Tt,1}(\xi,0,Q^{2})}^{2}
\end{equation}
where $b$ is a measure of the $t$-dependence of the DVCS amplitude and
currently expected to be nearly constant.  In Fig.~\reffig{sigma}
\begin{figure}
  \includegraphics[angle=-90,width=49\unitlength]{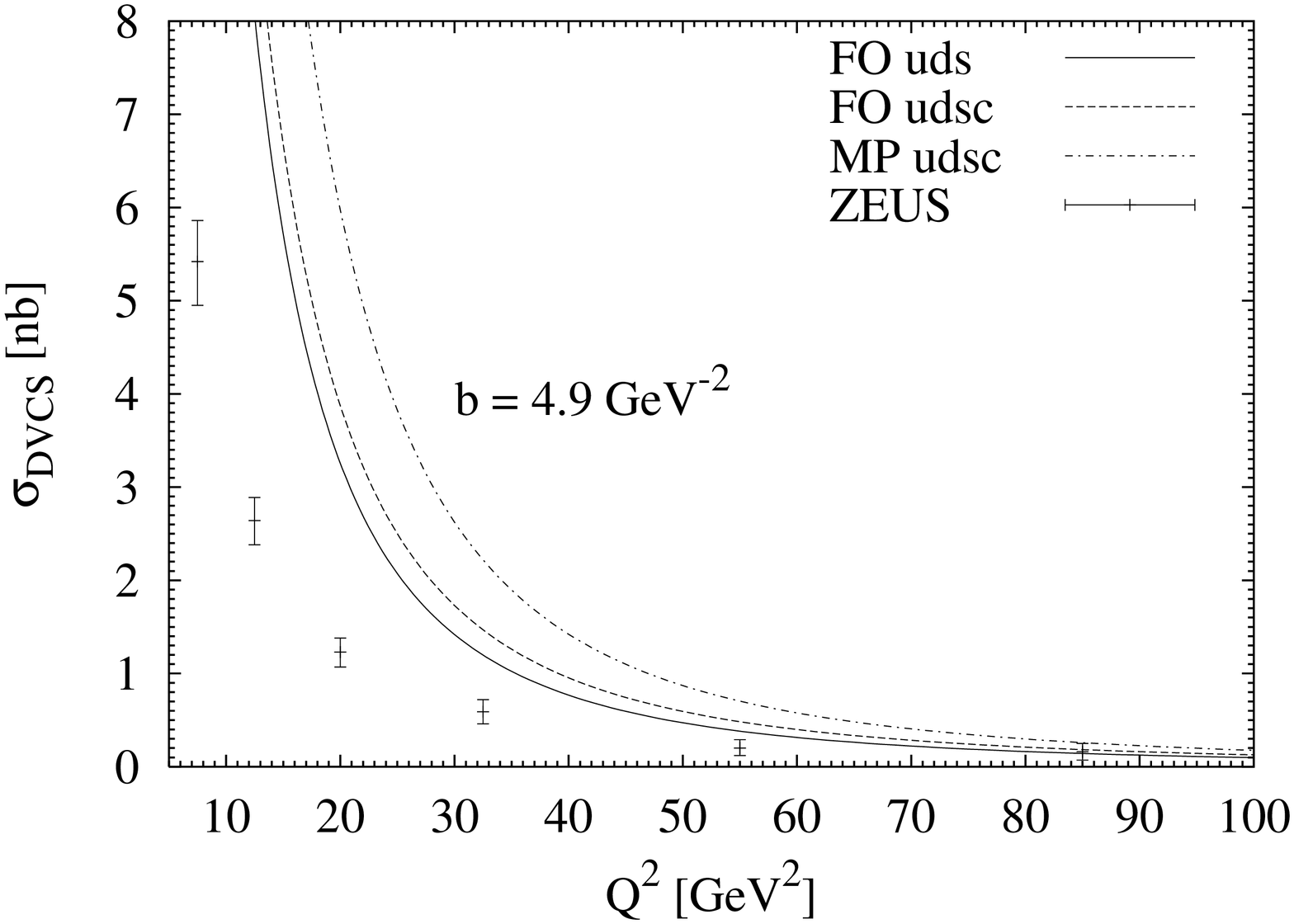}
  \hspace{\fill}
  \includegraphics[angle=-90,width=49\unitlength]{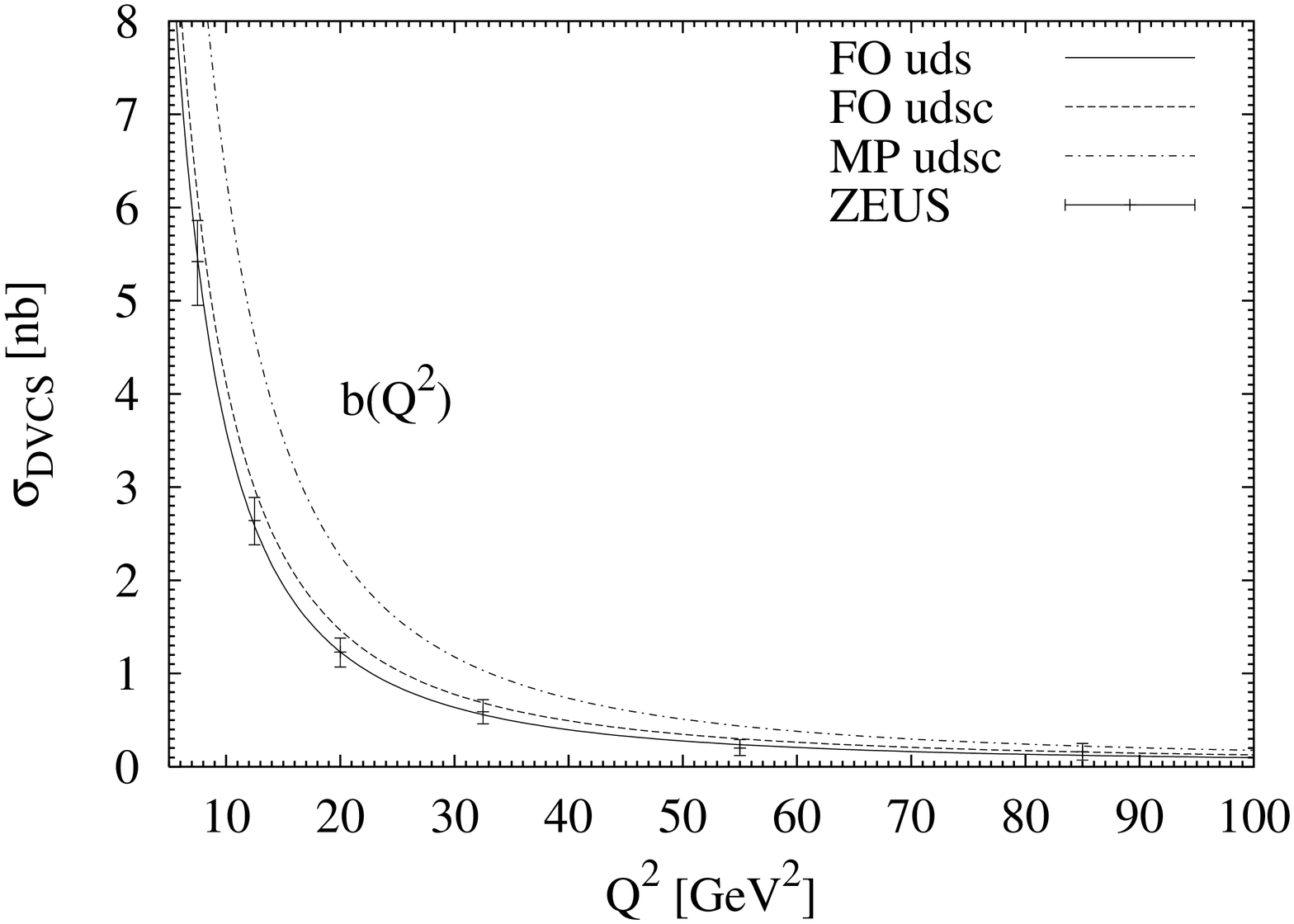}
  \caption{The DVCS cross section as a function of $Q^{2}$ at fixed
    $W=89\,\GeV$ for three light flavors (solid line), with the fixed
    order charm contribution (dashed line), and with the \massless\
    charm (dashed-dotted line), using a constant $b[\GeV^{-2}]=4.9$ as
    well as a strongly $Q^{2}$-dependent
    $b(Q^{2})[\GeV^{-2}]=4.9-5\log(Q^{2}[\GeV]/100)$.  The error bars
    denote the quadratic sums of the statistical and the systematic
    uncertainties of the ZEUS data~\cite{Chekanov:2003ya}.}
  \labelfig{sigma}
\end{figure}
the results for the DVCS cross section together with the ZEUS data at
fixed $W=\sqrt{\shat}=89\,\GeV$~\cite{Chekanov:2003ya} are shown for a
constant $b[\GeV^{-2}]=4.9$ as it has been used by the ZEUS
collaboration and additionally for a highly $Q^{2}$ dependent
$b(Q^{2})[\GeV^{-2}]=4.9-5\log(Q^{2}[\GeV]/100)$.  The constant $b$
shows the aforementioned difficulties by overshooting the data
considerably that is basically in line with previous leading order
predictions~\cite{Freund:2001hd}.  Only recently it was found that a
``$k$-delta ansatz''~\cite{Radyushkin:1998es} with $k=0$, i.~e.\ a
narrow $y$-dependence for the double distribution $F(x,y)$, can be used
to describe the data~\cite{Freund:2003qs}. However, such an ansatz is
considered not realistic since it does not fulfill a certain symmetry
constraint for double
distributions~\cite{Mankiewicz:1998uy,Radyushkin:1998es}.

The simple highly $Q^{2}$ dependent $b$-fit that yields a good agreement
of the light flavor result with the data should merely give an
impression of the relevance of charm contributions.  The fixed order
charm contributions increase the DVCS cross section by $10\%$ for the
lowest $Q^{2}$ shown and up to $30\%$ for large $Q^{2}$ which is beyond
the relative uncertainties of the first four data points.  The
\massless\ charm GPD leads to significantly larger increases in the
range of $60\ldots90\%$ which excessively overestimates the physical PGS
subprocess.  To check for the reliability of this result the CTEQ4
forward PDFs~\cite{Lai:1997mg} have been used alternatively that give
minor modifications less than $10\%$ apart from a reduction of up to
$15\%$ around $Q^{2}\sim10\,\GeV^{2}$ that can be traced back to the
higher charm mass $m_{c}^{\text{CTEQ}}=1.6\,\GeV$ and the
correspondingly delayed generation of charm.  Finally it should be noted
that a change of the factorization scale
$\mu_{h}\to\sqrt{4\mhq^{2}+Q^{2}}$ changes the fixed order results below
the percent level.
%
%
\section{Conclusions and outlook}\labelsec{theend}
The contributions of heavy quarks, especially of the charm quark, to
deeply virtual Compton scattering have been adequately analyzed by fixed
order perturbation theory via photon-gluon scattering.  The use of
intrinsic \massless\ generalized heavy quark distributions overestimates
the DVCS cross section at small skewedness $\lesssim{}10^{-3}$.  In
principle this may be absorbed in models for GPDs, which, however, is
unreasonable since the fixed order calculation correctly describes at
least the threshold region.  At large scales $Q^{2}$ both treatments
cannot be distinguished because of present experimental uncertainties.

The rough estimates indicate that heavy quark mass effects are relevant
for small skewedness $\lesssim10^{-3}$ in particular at low scales with
respect to experimental uncertainties.  Nevertheless this should be
confirmed with ``realistic'' GPD models and using the full DVCS
amplitude, possibly with higher twist corrections.  For values of the
skewedness $\sim{}0.1$, characteristic for fixed-target experiments,
contributions of heavy quarks can safely be neglected.  The massive
analytical results for the photon helicity flip form factors of the DVCS
amplitude have not been utilized but can be trivially included in future
analyses.  In this case, the massive description for heavy quarks
should always be used as it smoothly reproduces a \massless\ parton
picture at high scales.

To prove the perturbative stability of the PGS model the next-to-lowest
order corrections to the massive hard scattering coefficients have to be
calculated which are necessary for a consistent next-to-leading order
analysis with heavy quarks correctly included.
%
%
\section*{Acknowledgments}
I am indebted to M.\ Gl\"{u}ck and E.\ Reya for proposing this
investigation as well as for helpful discussions and suggestions.  I
would also like to thank I.\ Schienbein for useful conversations and
carefully reading the manuscript.  This work has been supported in part
by the ``Bundesministerium f\"{u}r Bildung und Forschung'', Berlin/Bonn.
%
%

\end{document}